\documentclass[10pt,aps,prc,twocolumn,showpacs,showkeys,nofootinbib]{revtex4-1}
\usepackage{graphicx}
\usepackage{amsmath,amsfonts,mathrsfs}
\usepackage{colordvi}
\newcommand{\nl}{\nonumber\\ }
\newcommand{\pd}{\partial}

\newcommand{\vc}[1]{\vec{#1}}

\newcommand{\vU}{U}

\newcommand{\vD}{\Delta}

\newcommand{\Elab}{E_{\rm lab}}
\newcommand{\agev}{A\text{GeV}}
\newcommand{\rmd}{{\mathrm d}}
\newcommand{\rmid}{{\rm id}}

\newcommand{\Tfrz}{T_{\rm f.o.}}

\def\lsim{\mathrel{\rlap{
\lower4pt\hbox{\hskip-3pt$\sim$}}
    \raise1pt\hbox{$<$}}}     
\def\gsim{\mathrel{\rlap{
\lower4pt\hbox{\hskip-3pt$\sim$}}
    \raise1pt\hbox{$>$}}}     

\DeclareMathOperator{\Div}{div}

\begin{document}

\title{Constraints on shear stress tensor in viscous relativistic hydrodynamics}

\author{A. S. Khvorostukhin}
\email{hvorost@theor.jinr.ru}
\affiliation{Joint Institute for Nuclear Research, RU-141980 Dubna, Russia}
\affiliation{Institute of Applied Physics, Moldova Academy of Science, MD-2028 Kishineu, Moldova}

\author{E. E. Kolomeitsev}
\affiliation{Matej Bel University, SK-97401 Banska Bystrica, Slovakia}
\affiliation{Joint Institute for Nuclear Research, RU-141980 Dubna, Russia}

\author{V. D. Toneev}
\affiliation{Joint Institute for Nuclear Research, RU-141980 Dubna, Russia}

\begin{abstract}
We extend our hybrid model HydHSD by taking into account shear viscosity within the Israel-Stewart hydrodynamics. The influence of different forms of $\pi^{\mu\nu}$ constraints on observables is analyzed. We show that the form of the corresponding condition plays an important role for the sensitivity of viscous hydrodynamics to the ratio of shear viscosity to the entropy density, $\eta/s$. It is shown that the constraint used in the vHLLE model, results in most sensitivity of rapidity distributions and transverse momentum spectra to a change of the $\eta/s$ ratio; however, their applicability for large values of $\eta/s$ is doubtful. On the contrary, the strict constraints from \cite{MNR2010} are very strong but most established. We also found that $\eta/s$ as a function of the collision energy probably has an extremum at $\Elab=10.7\agev$.  However, we obtain that any considered condition does not allow to reproduce simultaneously pion and proton experimental data within our model.
\end{abstract}

\maketitle
\section{Introduction}

Hydrodynamics is a powerful phenomenological tool having a variety of wonderful properties. It allows one to take easily into account collective effects and the equation of state (EoS) of studied matter which cannot be completely described by microscopic models. Application of hydrodynamics to theoretical description of high-energy nuclear collisions  has been started with Landau's original work \cite{La53}. The actual status and successful story of hydrodynamics approach in ultra-relativistic heavy-ion collision theory is reflected in review articles \cite{KH03,HS13,GJS13,JH15,DKK16,FHS17}.

A problem of heavy ion collision modeling is that hydrodynamics applicability conditions are violated at the early and final stages of nuclear interaction. The main condition assumes that the mean free path of quasiparticles in a system has to be smaller than the system size. It is clear that this condition is not satisfied in dilute matter at the beginning and the end of a collision when medium is far from the local equilibrium.

One way to get around the mentioned problem is to construct a hybrid model. Within hybrid models, one of which we developed in~\cite{HYDHSD2015}, the initial conditions for hydrodynamic equations, i.e. space distributions of the energy density, charge density, and velocity field, are calculated using the kinetic model.

The previous version of our HydHSD hybrid model~\cite{HYDHSD2015} includes ideal hydrodynamics as a part. More realistic calculations of heavy-ion collisions at relativistic energies need to take into account  non-zero viscosity of QCD matter which is the aim of this paper. Here we study rapidity spectra and transverse momentum distributions of hadrons produced in relativistic  nuclear collisions in the range below the top SPS energy, $\Elab \le 160\ \agev$, in terms of our hybrid model.

The article is organized as follows. We start with the description of the set of viscous hydrodynamic equations in Sec.~\ref{hydroeq_sec} and how it is solved numerically, see Sec.~\ref{numerical_sec}. Sec.~\ref{initial_sec} is devoted to obtaining the initial conditions. In Sec.~\ref{observable_sec}, the particlization procedure used for observable calculation is shortly formulated. Some words about the EoS can be found in Sec.~\ref{eos_sec}. We continue the consideration in Sec.~\ref{param_depend} where  we discussed how our model depends on the parameters. A special attention is paid to the constraints on the shear stress tensor, see Sec.~\ref{VHLLEMUSIC}. Our final results are presented in Sec.~\ref{fitsection}.
Technical details of our numerical algorithm are given in Appendiсes.

\section{The Model}

\subsection{Equations of viscous hydrodynamics}
\label{hydroeq_sec}
The system undergoing hydrodynamic evolution is described by the set of equations~\cite{hydroabout}
\begin{subequations}
\begin{align}
\partial_\mu T^{\mu\nu}&=0,
\label{hydrobase-T}\\
\partial_\mu J^\mu &=0,
\label{hydrobase-J}
\end{align}
\label{hydrobase}
\end{subequations}
involving an energy-momentum tensor $T^{\mu\nu}$ and a baryon current $J^\mu$, and representing the conservation laws of the total energy, momentum, and baryon charge.
Here and below we will use the Cartesian coordinates. In the general case of a non-ideal fluid when dissipation processes are possible, the energy-momentum tensor and the baryon current can be cast in the form~\cite{hydroabout}
\begin{align}
T^{\mu\nu}&=T_{\rm id}^{\mu\nu} -\Pi\Delta^{\mu\nu}+\pi^{\mu\nu}, \quad
J^\mu=n\, u^\mu + V^\mu\,,
\label{Tmunu}\\
T_{\rm id}^{\mu\nu}&=\varepsilon\, u^\mu u^\nu-P\Delta^{\mu\nu}\,,
\label{T-ideal} \\
\Delta^{\mu\nu}&=g^{\mu\nu}-u^\mu u^\nu\,,
\label{Deltadef}
\end{align}
where $T_{\rm id}^{\mu\nu}$  is the ideal part of the energy-momentum tensor, $\varepsilon$, $n$, and $P$ are the energy density,  the baryon density, and the pressure in the local reference frame (LRF), $g^{\mu\nu}={\rm diag}(1,-1,-1,-1)$ is the metric tensor. The full energy-momentum tensor contains additional terms: the bulk pressure $\Pi$, the shear stress tensor $\pi^{\mu\nu}$, and the baryon (charge) diffusion current  $V^\mu$. The 4-velocity $u^\mu$ is defined here as an eigenvector of the full energy-density tensor
$T^\mu_\lambda\,u^\lambda=\varepsilon\, u^\mu$ (the Landau definition). It is normalized as $u_\lambda u^\lambda=1$ and can be written as $u^\mu = \gamma(1,\vc{v})$ through the 3-velocity $\vc{v}$ and $\gamma=(1-v^2)^{-1/2}$. From this definition of the flow velocity it follows that $\pi^{\mu\nu}$ is a traceless symmetric tensor satisfying the orthogonality relations:
\begin{align}
\label{orthogonality}
u_\mu\pi^{\mu\nu}&=0,\quad \pi^{\mu\nu}=\pi^{\nu\mu},\quad \pi^\mu_\mu=0.
\end{align}

Equations (\ref{hydrobase}) have to be supplemented by the EoS $P=P(\varepsilon,n)$.
If one considers a perfect fluid and puts $\pi^{\mu\nu}=0$, $\Pi=0$, and $V^\mu=0$, then the system of equations becomes closed and can be solved with respect to $T^{0\nu}_{\rm id}$ and $J^0$ taken as independent variables. In the viscous case, however, we need some additional equations for $\pi^{\mu\nu}$, $\Pi$,  and $V^\mu$ which are now independent dynamical variables. Below, for simplicity, we neglect the heat flux, i.e., $V^\mu = 0$ is assumed.  Note that in this case the Landau and Eckart frames coincide.

Studies performed in~\cite{DMNR2012} show that there can be infinitely many choices for the explicit form and coefficients in the equations of motion for $\pi^{\mu\nu}$ and $\Pi$. In this work, we follow the original Israel-Stewart framework~\cite{IS}, in which all viscous terms of the second order in gradients are suppressed. Additional quantities are governed by the relaxation-type equations
\begin{align}
(u^\lambda\partial_\lambda)\,\Pi &=-\frac{\Pi+\zeta\theta}{\tau_\Pi},
\quad \theta= \partial_\lambda u^\lambda\,,
\nonumber\\
(u^\lambda\partial_\lambda)\,\pi^{\mu\nu}&=-\frac{\pi^{\mu\nu}-\eta W^{\mu\nu}}{\tau_\pi}\, ,
\label{ISeqs}\\
W^{\mu\nu} &= \Delta^{\mu\lambda}\partial_\lambda u^\nu
+ \Delta^{\nu\lambda}\partial_\lambda u^\mu-\frac23\,\Delta^{\mu\nu}\, \theta\,,
\nonumber
\end{align}
where $\tau_\Pi$ and  $\tau_\pi$ are the relaxation times for the bulk pressure and the shear stress tensor while $\zeta$ and $\eta$ are the bulk and shear viscosity, respectively.      For vanishing relaxation times, $\tau_\pi=\tau_\Pi=0$, Eqs.~(\ref{ISeqs})
lock viscous terms $\Pi$ and $\pi^{\mu\nu}$ to their first-order values $-\zeta\theta$ and $\eta W^{\mu\nu}$, respectively, so that replacing them in Eqs.~(\ref{hydrobase}) and (\ref{Tmunu}) we recover the well-known Navier-Stokes formulaes.

Below in this work, we will neglect the bulk viscosity and put $\zeta=0$.
Then the system of hydrodynamic equations is closed by the expressions for $\tau_\pi$ and $\eta$, which in principle have to be calculated consistently with the EoS. However, in our calculations we use simplified relations~\cite{MHHN2014}
\begin{align}
\eta = k_\eta\,s,\quad\tau_\pi = \frac{5\eta}{\varepsilon+P}\,,
\label{eta-tau-def}
\end{align}
where $k_\eta={\rm const}$ and the entropy density $s=s(\varepsilon, n)$ is given by the EoS.

Finally, we quote the expression for the entropy. We need only zero component of the entropy 4-vector, $s^\mu$, which reads for a cell as
\begin{align}
\label{s0def}
s^0&=\Big(s-\frac{\tau_\pi}{4T\eta}\,\pi^{\mu\nu}\pi_{\mu\nu}
-\frac{\tau_\Pi}{2T\zeta}\,\Pi^2\Big)\,u^0
\end{align}
with the temperature $T=T(\varepsilon, n)$ given by EoS. Then the total  entropy is the sum of $s^0$ over all cells multiplied by the cell volume. The dissipative part in Eq.~(\ref{s0def}) can be larger than the first term, $s$, in some cells. This can happen due to numerical errors and because viscous hydrodynamics, being applied to heavy ion collisions, works  at the edge of its applicability range. To get rid off these artifacts,  we exclude cells with temperatures $T<50$\,MeV in calculations of the total entropy.

\subsection{Numerical scheme}
\label{numerical_sec}

For the numerical implementation, we first of all have to specify independent variables in the equations of motion (\ref{hydrobase}) and (\ref{ISeqs}). In viscous hydrodynamical codes, one usually takes the $J^0$ and $T^{0\nu}$ components of the energy-momentum tensor. In this case, the reconstruction of the LRF quantities such as energy and baryon densities and  the 3-velocity of the fluid cell becomes a complicated problem~\cite{MNR2010,Muronga07}. Instead, we will use the components of the ideal-fluid tensor $T^{0\nu}_{\rm id}$ as independent variables, which allows us to apply relations (\ref{recover}), (\ref{recover-v}), and (\ref{recover-v-dir}) without a problem.
Then evolution of $T^{0\nu}_{\rm id}$ is described by the equation
\begin{align}
\label{hydrodecomposition}
    \pd_\mu T_{\rmid}^{\mu\nu}&=-\pd_\mu\pi^{\mu\nu}\,,
\end{align}
which is just a rewriting of Eq.~(\ref{hydrobase-T}).

As follows from Eq.~(\ref{orthogonality}), only five components of the $\pi^{\mu\nu}$ tensor are independent.
The other can be reconstructed if the cell velocity is known. However, for some choices of this five component set, the reconstructed components can contain a singularity if an element of the vector $\vc{v}$ vanishes~\cite{MNR2010}. We select $\pi^{yy}$, $\pi^{zz}$, $\pi^{xy}$, $\pi^{xz}$, and $\pi^{yz}$ as independent ones in our implementation of the algorithm.
As one can see from Eq.~(\ref{recover-pi}),  a singularity is absent for the such choice if $v<1$.

Equations~(\ref{hydrobase-J}) and (\ref{hydrodecomposition})  can be rewritten in the form
\begin{align}
\label{shastaform}
\pd_t\vc{U}_{\rm cons}&+\sum_i\pd_i(v_i\vc{U}_{\rm cons})=\vc{S}_{\rm cons}
\end{align}
where we introduced a 5-dimensional vector for generalized densities $$\vc{U}_{\rm cons}=(J^0,T_{\rm id}^{00}, T_{\rm id}^{0x}, T_{\rm id}^{0y}, T_{\rm id}^{0z})^{\rm T}$$
and the corresponding sources $\vc{S}_{\rm cons}$, see Appendix~\ref{app:shasta}.
This set of equations is solved numerically by means of the SHASTA (the SHarp and Smooth Transport Algorithm) algorithm~\cite{SHASTA,SHASTARischke}. In this article, we follow the numerical scheme outlined in Section 4.2 of Ref.~\cite{MNR2010} extending it to $3+1$ dimensions. The corresponding formulae are collected in Appendix~\ref{app:shasta}.

In principle, the relaxation equations~(\ref{ISeqs}) can also be solved in similar way but as shown in \cite{MHHN2014}, the  algorithm becomes more stable if one uses a simple centered second-order differences scheme  for spatial gradients on the left-hand
side of Eqs. (\ref{ISeqs}). So, the shear stress tensor components, $\vc{\pi}=(\pi^{xy}, \pi^{xz}, \pi^{yz}, \pi^{yy}, \pi^{zz})^{\rm T}$, are propagated  according to the described simple scheme.

In some cells the relaxation time $\tau_\pi$ given by Eq.~(\ref{eta-tau-def}) may become smaller than the calculation time step. Then, following the idea from Section 3.2 of Ref.~\cite{KHB2013}, we evolve $\pi^{\mu\nu}$ using the formal solution of Eq.~(\ref{ISeqs})\footnote{Eq. (\ref{pi_formal}) is applied before the antidiffusion step.}
\begin{align}
\label{pi_formal}
\pi^{\mu\nu}(t_{n+1})&=[\pi^{\mu\nu}(t_n)-\eta W^{\mu\nu}]e^{-\Delta t/(\gamma\tau_\pi)}+\eta W^{\mu\nu}.
\end{align}

It is important that at each calculation step we have to ensure the applicability of the hydrodynamic equations. We have to be sure that viscous effect are kept only as corrections to the ideal fluid energy-momentum tensor. Therefore, at each time step in each cell we calculate the ratio

\begin{align}
q=q_{\rm S}=\max_{\mu,\nu}\frac{|\pi^{\mu\nu}|}{|T_{\rm id}^{\mu\nu}|},\quad \mbox{(S-cond.)}
\label{q-def-S}
\end{align}
and verify the fulfillment of the condition~\cite{MNR2010}
\begin{align}
q<C,
\label{piconstrain}
\end{align}
where $C$ is a predefined positive constant, $C<1$. If the opposite occurs we rescale the shear stress tensor as
\begin{align}
\pi^{\mu\nu}\to \pi^{\mu\nu}_{\rm corr}=\pi^{\mu\nu}\frac{C}{q}\,.
\label{rescale}
\end{align}
Such a rescaling prescription is frequently used in the literature~\cite{KHB2013,KHPB,VISHNU,MUSIC}; however, there are differences in how the quantity $q$ is evaluated.
This aspect will we considered in detail in Section~\ref{VHLLEMUSIC}.
The condition (\ref{piconstrain}) evaluated with $q$ from Eq.~(\ref{q-def-S}) will be denoted as the strict (S-) condition.
By default we use $C=0.3$.

Heavy ion collisions at relativistic energies are believed to produce a deconfined, strongly coupled quark-gluon plasma (QGP)~\cite{Shuryak}. In the initial stages of the collision, during which the QGP is produced, the system is surely far from equilibrium and cannot be described by hydrodynamics. However, modeling based on near-ideal hydrodynamics strongly suggests that a hydrodynamic treatment becomes applicable rather quickly, on times approximately less than 0.1\,fm/c after the collision event at RHIC energies~\cite{Heinz2005}. Some aspects of this process involving strongly coupled dynamics can be studied in theories, which describe non-Abelian plasmas similar to the QGP, and which possess dual gravitational descriptions, the  best known example being $N= 4$ supersymmetric Yang-Mills (SYM) theory~\cite{SYM}. Using this gauge/gravity duality, it is possible to study how quickly a far-from-equilibrium strongly-coupled non-Abelian plasma relaxes to a near-equilibrium regime, in which a hydrodynamic description is getting accurate, and to estimate energy and entropy of the formed system\footnote{It was noted~\cite{Romatschke2017,Attems2017} that the system created in high energy nuclear collisions reaches or at least comes close to equilibrium. In particular, it was realized that because of the expansion of the matter into the vacuum, the system would cool and thus freeze into a hadronic gas quickly. Thus, it became apparent that a fluid dynamic approximation to the system dynamics had to start early, on a time-scale of $\tau\sim 1$ fm/c or less.}.

Results obtained in \cite{holography, Romatschke2017} confirm that the 2nd order hydrodynamics is applicable when
\begin{align}
\label{neareqcondition}
  |\pi^{\mu\nu}|<|T^{\mu\nu}_{\rm id}|.
\end{align}
However there is a misunderstanding regarding these papers since authors  	assiduous underline that neither local near-equilibrium nor near isotropy are required for hydrodynamics applicability. So we paid attention that the condition (\ref{neareqcondition}) allows a large anisotropy of the pressure and a  state which is quite far from equilibrium. But shear pressure satisfying the Eq. (\ref{neareqcondition}) or weaker one is often named in the literature as giving sufficiently small corrections to the equilibrium quantities, see, e.g., [1, 17]. It is apparent that it not completely clear statement since viscous contribution are large at least when $|\pi^{\mu\nu}|\gtrsim\frac12|T^{\mu\nu}_{\rm id}|$. Below when we speak about a close/near to equilibrium state, we mean vicinity defined by Eq. (\ref{neareqcondition}).

Let us also mention that one has to be careful with conclusions of \cite{Attems2017} since as is seen from Fig. 5 of the cited reference, $P_T\neq0$ when $P_{eq}=0$. It is not the case of usual matter.

\begin{figure}
\centering
\includegraphics[width=5.5cm]{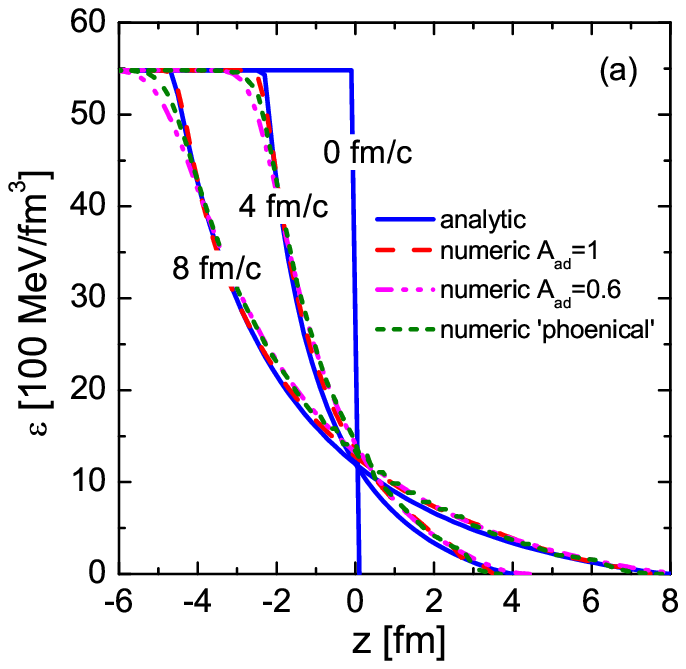} \\
\includegraphics[width=5.5cm]{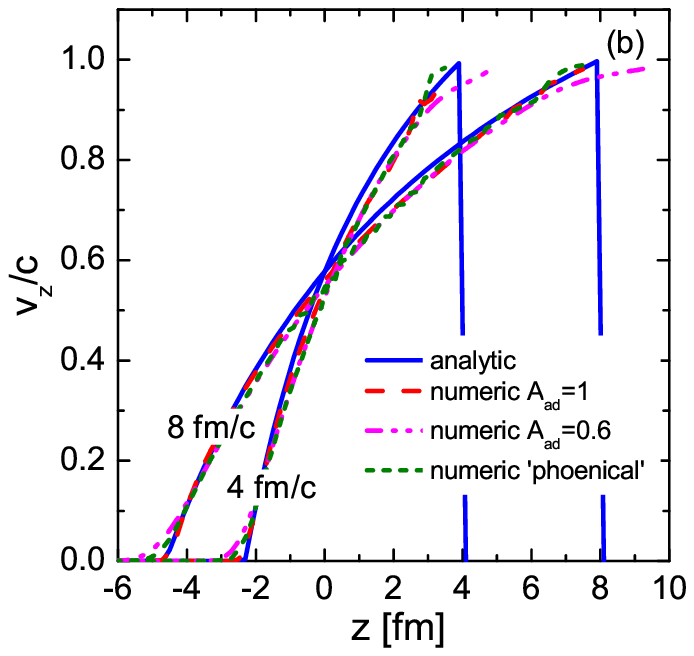}
\caption{A test of the hydrodynamical code on the solution of the (1+1)-dimensional Riemann problem. Energy density (a) and velocity profiles (b)  calculated at various moments of time and in comparison with analytic solutions~\cite{Skokov-Toneev07} are shown.
Short-dashed lines are the results of the application of the phoenical version of the SHASTA. Dashed and dash-dotted lines are calculated within the explicit SHASTA for various values of the mask coefficient $A_{\rm ad}$ controlling anti-diffusion.} \label{analyttest}
\end{figure}
To verify our hydrodynamical code, we performed a test similar to that proposed in Ref.~\cite{MNR2010}, namely, we solved numerically the (1+1)-dimensional Riemann problem for two states with constant pressure (or energy density since $p=\varepsilon/3$) equal to $p_0$ on one side and to 0 (vacuum) on the other side separated by a membrane located at $x = 0$. The evolution of energy density and velocity profiles is presented in Figs.~\ref{analyttest}a and~\ref{analyttest}b, respectively. We used here $\Delta x = 0.2$\,fm and put very small shear viscosity, $\eta/s=0.01$ to simulate numerically a viscous free flow. We see a good agreement of numerical solutions with analytical ones.

\begin{figure}\centering
\includegraphics[width=5.5cm]{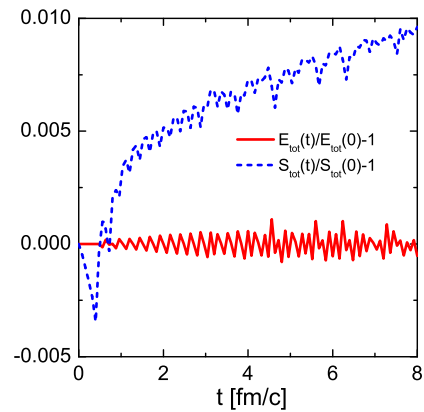}
\caption{Time dependence of $E_{\rm tot}(t)/E_{\rm tot}(t_0)-1$ ( the solid line) and $S_{\rm tot}(t)/S_{\rm tot}(t_0)-1$ (dashed one)  obtained for the numerical solution of the (1+1)-dimensional Riemann problem with $\eta/s=0.01$ and $A_{\rm ad}=1.0$. \label{es-evol-test}}
\end{figure}

However, the price for imposing the constraint~(\ref{piconstrain}) by hand is that the total energy and total entropy of the system become non-monotonic functions of time, see Fig.~\ref{es-evol-test}. The total energy fluctuates around a constant value while the entropy oscillates about a slowly increasing average value. The reason of such behaviour is the rescaling procedure (\ref{rescale}). Note that the total energy is a constant at the beginning, since we start with $\pi^{\mu\nu}=0$ and no $\pi$-rescaling is needed for the first several time steps. Small initial entropy decreasing is just due to numerical inaccuracy.

We also checked our algorithm by 1D boost-invariant expansion \cite{ECHOQGP}.

\begin{figure}
\centering
\includegraphics[width=6cm]{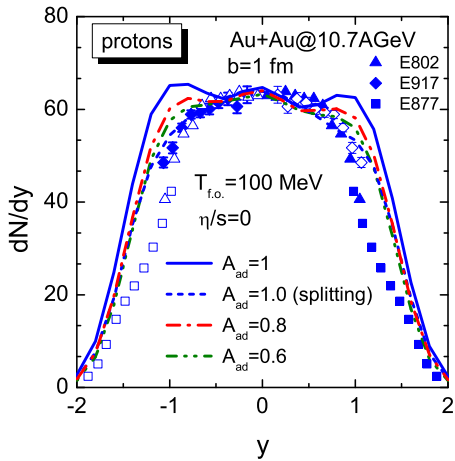}
\caption{The comparison of ideal proton rapidity distributions  a the freeze-out temperature $\Tfrz=100$\,MeV for $\Elab=10.7\,\agev$, calculated within the 'phoenical' version  of SHASTA with 3D-splitting (dash-dotted line) and 'explicit' complete 3D one, see \cite{MNR2010,NDHMR12}.
Experimental points are from Refs.~\cite{PRC57,PRC62,PRL86}. }\label{dNdy_E10_T100}
\end{figure}

Let us remind that in our previous work~\cite{HYDHSD2015} we solved equations of ideal hydrodynamics by the `phoenical' version of SHASTA~\cite{SHASTARischke} and used the operator-splitting method to treat three-dimensional operators. Our new code implements the `explicit' SHASTA~\cite{SHASTARischke} and straight away solves the 3D problem.  So, as one additional test of our new code, we can compare proton rapidity distributions calculated within ideal hydrodynamics by old `splitting' and new versions of SHASTA. The corresponding results for the Au+Au collision at $\Elab=10.7\,\agev$ with the freeze-out temperature $\Tfrz=100$ MeV are shown in Fig.~\ref{dNdy_E10_T100}. One sees that if we use the default value of the so-called mask coefficient\footnote{The algorithm~\cite{MNR2010} allows one to use different mask coefficients for every direction, $x$, $y$, and $z$, but for simplicity we take one value for all axes.}, $A_{\rm ad}=1$, the rapidity distributions essentially differ from each other at intermediate  rapidity values where the new code leads to humps.
We need to note that such differences disappear with growing $\Tfrz$ and are caused by too large antidiffusion in peripheral cells.     The latter conclusion is confirmed by Fig.~\ref{dNdy_E10_T100} where the humps are suppressed if $A_{\rm ad}$ decreases until $A_{\rm ad}=0.6-0.5$.

Large antidiffusion in the "explicit" SHASTA was early mentioned in~\cite{NDHMR12} where authors proposed to set $A_{\rm ad}$ to be proportional to $1/((k/\varepsilon)^2+1)$, where $k$ is some small constant of order $10^{-5}\,{\rm GeV/fm^3}$. In this way, $A_{\rm ad}$ goes smoothly to zero near the boundaries of the grid, i.e. we increase the amount of numerical diffusion in that region~\cite{NDHMR12}. However, such choice does not affect the solution while we have a more complicated problem and need to suppress humps.

For our calculations, we take $A_{\rm ad}=0.6$ which noticeably decrease the proton yield in the fragmentation region, see Fig.~\ref{dNdy_E10_T100}.

\subsection{Initialization of hydrodynamic evolution}
\label{initial_sec}

The differential equations of hydrodynamics must be supplemented by appropriate initial conditions. As was noted in the previous Section B, in the initial stage a deconfined strongly coupled QGP phase may be formed. Its non-equilibrium evolution may be described using gauge-gravity duality. In hybrid models these conditions are usually deduced from results of calculations in some kinetic model. It would also allow one for an event-by-event analysis of collisions. In our approach we use the Hadron String Dynamics (HSD) model~\cite{HSD-1,HSD-2,HSD-3} which is very successful in the description of experimental data in the considered energy range.~\footnote{Particularly we use version 1.0 of the Parton-Hadron String Dynamics model with the switched off partonic option}. In order to obtain relatively smooth initial distributions of the energy-momentum density and the baryon number, one can either perform averaging over many collision events or smear particles for a selected event in space with the help of a Gaussian distribution, for example~\cite{Oliin-Petersen15}. In our approach we calculate the quantities
\begin{align}
T^{\mu\nu}_{\rm init} (\vec{r}) &=\overline{
\sum_a\frac{p_a^\mu\,p_a^\nu}{p_a^0} K(\vec{r}-\vec{r}_a)
}\,,
\nonumber\\
J^\mu_{\rm init}(\vec{r}) &=\overline{
\sum_a\frac{p_a^\mu}{p_a^0} K(\vec{r}-\vec{r}_a)
}\,,
\label{TJ-init}
\end{align}
where the bar stands for the event averaging and the sum runs over particles at the positions $\vec{r}_a$, $K(\vec{r})$ is the smoothing function which in our case performs just averaging in the volume element, $\Delta V$
\begin{align}
K(\vec{r})=\left\{\begin{array}{ll}
1/\Delta V &,\,\, \vec{r}\in\Delta V\\
0 &, \,\, \vec{r}\notin\Delta V
\end{array}
\right.
\,.
\end{align}
Assuming the $T_{\rm init}^{\mu\nu}$ structure as for an ideal fluid, see Eq.~(\ref{T-ideal}) and the absence of the baryon diffusion current, $V^\mu=0$, we obtain from quantities~(\ref{TJ-init}) the initial energy density, $\epsilon_{{\rm init}}$, and the baryon density, $n_{{\rm init}}$, in a fluid cell and the cell velocity, $\vec{v}$, with the help of relations (\ref{recover}), (\ref{recover-v}), and (\ref{recover-v-dir}). Now we can evaluate the initial entropy and other thermodynamical quantities in each cell using the equation of state, e.g., $s_{{\rm init}}=s(\epsilon_{{\rm init}}, n_{{\rm init}})$\,, cf. Eq. (\ref{s0def}). The initial total entropy and the baryon number of the system is calculated as
\begin{align}
S_{\rm init}=\sum_{\rm cell} \frac{s^0_{{\rm init}}}{\sqrt{1-\vec{v}^2}}\,,
\nonumber\\
N_{\rm init}=\sum_{\rm cell} \frac{n_{{\rm init}}}{\sqrt{1-\vec{v}^2}}\,.
\label{SN-init}
\end{align}
The advantage of this method is that it conserves energy and momentum. However, this procedure supports switching only to an ideal fluid, neglecting viscous corrections. Therefore, at the beginning of the hydrodynamical stage all components of the shear-stress tensor are initialized with zero values. That is found to be a useful approximation in the literature.

A transition from a kinetic to a hydrodynamic regime occurs at an instant $t_{\rm start}$. We assume that at this moment the system is close to equilibrium and the ratio
of the entropy to the baryon number $S(t)/N(t)$ ceases changing, see Fig.~1 in Ref.~\cite{HYDHSD2015}.

Below we consider the following heavy-ion collisions:
Au$+$Au collisions for AGS energies at $\Elab = 6$ and 10.7\,$\agev$, and
Pb$+$Pb collisions for SPS energies at $\Elab = 40$, 80, and 158\,$\agev$.
All calculations are performed for the impact parameter $b=1$\,fm.
The corresponding transition times from HSD to hydrodynamics, $t_{\rm start}$ are shown in Table~\ref{tabl1} together with the corresponding starting entropy. These times are not changed comparing with our previous work~\cite{HYDHSD2015}.

\setlength{\tabcolsep}{5pt}
\begin{table}
\caption{Starting times of hydrodynamical calculations} \label{tabl1}
\begin{tabular}{lccccc}
\hline\hline
$\Elab$\,[\agev]          &6  & 10.7& 40 & 80 & 158\\
  \hline
$t_{\rm start}$\,[fm/$c$]&7.9&7.18 &4.57&3.8 & 3.01\\
\hline
$S_{\rm start}\times 10^{-3}$
  &  3.5 &4.4   &  7.1   &  8.7    & 10.6 \\
\hline\hline
\end{tabular}
\end{table}

\subsection{Particlization procedure and observables}
\label{observable_sec}

To convert fluids to particles, we realized a particlization procedure  according to the Cooper-Frye formulae~\cite{HYDHSD2015}:
\begin{align}
\label{CooperFrye}
E\frac{\rmd^3N_a}{\rmd p^3}&=\frac{g_a}{(2\pi)^3}\int\rmd\sigma_\mu  p^\mu f_a(x,p)\,, \end{align}
where $p^\mu=(E,\vc{p})$ is the particle 4-momentum, $f_a(x,p)$ represents the distribution function (Wigner function) of the particle of type `$a$' and $g_a$ is the corresponding spin-isospin degeneracy factor, $\rmd\sigma_\mu=n_\mu\rmd^3\sigma$ is an element of the space-time freeze-out hypersurface with the normal  $n_\mu$.
The freeze-out hypersurface, as in the previous work~\cite{HYDHSD2015}, is determined with the help of the CORNELIUS algorithm~\cite{Huovinen}.

In the ideal fluid case, the particle distribution function is given by usual
the Fermi/Bose distribution
\begin{align}
f_{a}^{(0)}(x,p) &= \frac{1}{e^{\beta(p^\nu u_\nu(x)-\mu_a(x))} \pm 1}\,,
\end{align}
where $\beta=1/T$ is the inverse local temperature, $\mu_a$ is the chemical potential of the particle of type $a$ (Recall that the Coulomb interaction is neglected and all particles within in a given isospin multiplet have the same chemical potential). The plus and minus signs correspond to fermions and bosons, respectively.
For viscous fluids, one has to  take into account the modification of the distribution function because of non-equilibrium viscous effects. The common way is to approximate the distribution function by the following expression~\cite{KHPB}:
\begin{align}
f_a(x,p)&=f_{a}^{(0)}(x,p)
\nonumber\\
&\times\left[1+(1\mp f_a^{(0)}(x,p))\frac{p_\mu p_\nu\pi^{\mu\nu}}{2T^2(\varepsilon+P)}\right]\,.
\label{viscdistr}
\end{align}

We use exactly the same methods of particle momentum generation~\cite{PSBBS,FastMC} as described in \cite{KHPB}. To use it, one has to convert $\pi^{\mu\nu}$ to the LRF. Due to the orthogonality relations~(\ref{orthogonality}) are explicitly fulfilled in our code, we have $\pi^{*0\nu}=0$ where the asterisk refers to the LRF.

Another difference, in comparison to the ideal hydrodynamics, is that to apply the rejection procedure, one has to know the upper limit of the viscous correction factor. So one assumes that the square brackets in Eq.~(\ref{viscdistr}) should not be larger than 2, since the viscous term has to be only a small correction and, definitively, cannot be larger than unity.
As a check we also make runs letting the square brackets to be as large as 3. The  difference in the final momentum distributions of protons and pions is very small.

To calculate the proton fraction among nucleons, we use isospin factor $1/2$ while for pions $1/3$.

After generating ``thermal'' contributions by Eq.~(\ref{CooperFrye}) and (\ref{viscdistr}), resonance decays are taken into account in the zero-width approximation.

\subsection{Equation of state}
\label{eos_sec}

The used EOS~\cite{SDM09} includes all known hadrons with masses up to 2\,GeV in the zero-width approximation.  The equation of state of hadron resonance gas at finite temperature and baryon density is calculated thermodynamically taking into account a density-dependent mean field that guarantees the nuclear matter saturation.

To account for mean-field effects, an effective potential $U = U(n)$ acting on a hadron is introduced. It depends only on the baryon density, $n$, and does not depend on momenta of interacting baryons. Then the baryon’s single-particle energy can be obtained simply by adding $U(n)$ to the kinetic energy. In this case, the partition function of the hadronic system can be calculated analytically~\cite{PRD01}. As the result, the following expressions for
thermodynamic functions of the hadron EoS can be written:
\begin{align}
P &= \sum_a P_a(T,\mu^*,\mu_S) + P_f (n),
\\
\varepsilon &= \sum_a\varepsilon_a(T,\mu^*,\mu_S) + \varepsilon_f (n),
\end{align}
where the effective baryon chemical potential, $\mu^*$, is obtained by the shift $\mu^* = \mu_B - U(n)$.
The ``field'' contributions (marked by index 'f') to the
energy density and pressure are found as
$$
\epsilon(n) = nU(n) - P_f (n) =\int_0^n dn_1 U(n_1).
$$
In this approach meson contributions are given  by ideal gas expressions.

The mean-field potential is parameterized in a line with the Skyrme approach as $U(n) =\alpha n/n_0+ \beta(n/n_0)^\gamma$, where $n_0$ is the saturation density of nuclear matter
and $\alpha, \beta, \gamma ={\rm const}$. In the following, we fix $\gamma= 7/6$
and choose the remaining parameters from the requirements $P(T=0,n_0) = 0,\ \varepsilon(T=0,n_0)/n_0 = E_b+m_N$ where the binding energy $E_b=-16$\,MeV and $n_0=0.15\,{\rm fm}^{-3}$.

For more details on the EOS, see Ref.~\cite{SDM09}.

In the present study, we refrain from additional tunings of the EOS and the initial state.
We try to explain experimental data using only hadronic EoS to find observables, which cannot be described by a simple refitting of hydrodynamical parameters. Another reason is that changing EoS gives the additional very flexible degree of freedom and, by our opinion, should be used when the model parameter space will be well investigated.

\section{Influence of model parameters on momentum spectra}
\label{param_depend}

In this section we consider how a variation of the hydrodynamic model parameters  can manifest itself in rapidity ($y$) distributions and transverse momentum ($m_{\rm T}$) spectra at  $y=0$ of protons and pions. To be specific, we consider Pb+Pb collisions at 40\,$\agev$.

\subsection{Shear viscosity}

\begin{figure}
\includegraphics[width=8.8cm]{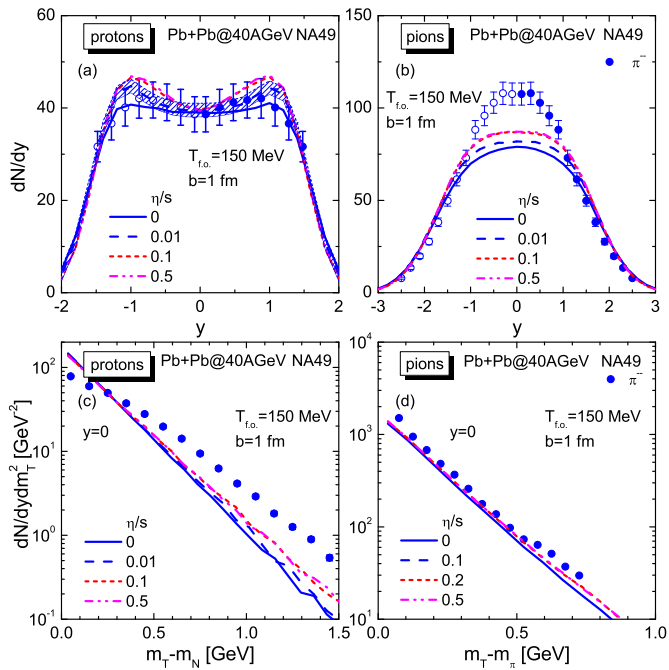}
\caption{Rapidity distributions and transverse momentum spectra for protons and pions produced in Pb+Pb collisions at 40\,$\agev$ in comparison with the calculations of the HydHSD model for various values of the $\eta/s$ ratio. The freeze-out temperature is fixed at $\Tfrz=160$\,MeV. Experimental points are taken from Refs.~\cite{SPSN1,SPSN2,SPSpiK}. The hatched area in panel (a) illustrate uncertainty in the number of produced particles induced by the Monte Carlo particlization procedure.
}
\label{fig:40AGeV-eta}
\end{figure}

First of all,  let us compare proton and pion $y$- and $m_{\rm T}$-distributions evaluated for viscous and ideal hydrodynamics. The results are collected in Fig.~\ref{fig:40AGeV-eta}. One expects that calculations with a very small value of $\eta/s =0.01$ have to be very close to ideal-hydro calculations. However, Figs.~\ref{fig:40AGeV-eta}a and~\ref{fig:40AGeV-eta}b
demonstrate that even such a small viscosity changes visibly the proton and pion rapidity distributions. One should however take into account the systematic uncertainty ($\sim\pm 1$ nucleons) in the particle number because of the Monte Carlo particlization procedure. Also the choice of the $A_{\rm ad}$ parameter influences the hump structure of the $dN_p/dy$ which becomes less pronounced and closer to the ideal hydrodynamics results for smaller values of $A_{\rm ad}$. Both mentioned error sources are further included in errors of fitted parameters, see Table \ref{tab:fit-parameters}, and do not influence on qualitative conclusions which we make below. For the $m_{\rm T}$ spectra the difference between calculations with $\eta/s=0$ and $\eta/s=0.01$ are small, as we see in panels (c) and (d) of  Fig.~\ref{fig:40AGeV-eta}.
An increase in the viscosity up to $\eta/s=0.1$ leads to sizable changes in the rapidity distributions, see the dashed lines in Figs.~\ref{fig:40AGeV-eta}a and~\ref{fig:40AGeV-eta}b. Particularly, the two-hump structure in the proton rapidity distribution becomes much more pronounced for the viscous case. It occurs because the shear viscosity slows the fireball longitudinal expansion and the fluid velocity which is reflected in the form of the $y$-distribution. Nonzero shear viscosity narrows the proton rapidity distributions and because of the
baryon number conservation, the narrowing leads to an increase of the hump height. At the same time, the height of proton distribution at mid-rapidity ($y=0$) is almost independent of $\eta/s$. Oppositely, the viscous corrections make the pion rapidity distributions higher than in the ideal case, see Fig.~\ref{fig:40AGeV-eta}b, as was anticipated  in Ref.~\cite{HYDHSD2015}.

The transverse momentum spectra of pions and protons show a very weak dependence on the $\eta/s$ value, see Figs.~\ref{fig:40AGeV-eta}c and~\ref{fig:40AGeV-eta}d, especially for pions. The inclusion of viscosity leads to a slight increase of slopes of the $m_{\rm T}$ spectra which makes  the spectra closer to the experimental data. It can be considered as an additional argument for the necessity of non-zero shear viscosity.

The slope of the calculated pion $m_{\rm T}$ spectrum roughly agrees with the experimental one, while for protons the calculated spectrum is too steep and underestimates the data for $m_{\rm T}-m_N>120$\,MeV.

Remarkably, in Fig.~\ref{fig:40AGeV-eta} we observe saturation of the viscosity effects with an increase in the $\eta/s$ ratio; indeed the lines calculated for $\eta/s=0.2$ and $0.5$ are barely distinguishable. This is because of the strict constraint on the $\pi^{\mu\nu}$ tensor (\ref{piconstrain}) with (\ref{q-def-S}), which we apply in our calculations, and the large gradients appearing in collisions at this energy. As we will see below, this saturation effect is specific for quite high energies and, for example, the sensitivity to $\eta/s$ is higher for collisions at $6\,\agev$, see Fig.~\ref{fig:6AGev-pi} below. However, this property together with the not too large increase of the pion rapidity distribution height leads to that the experimental data  for $\Elab=40\agev$ still cannot be reproduced.

\subsection{Freeze-out temperature}

\begin{figure}[h]
\includegraphics[width=8.8cm]{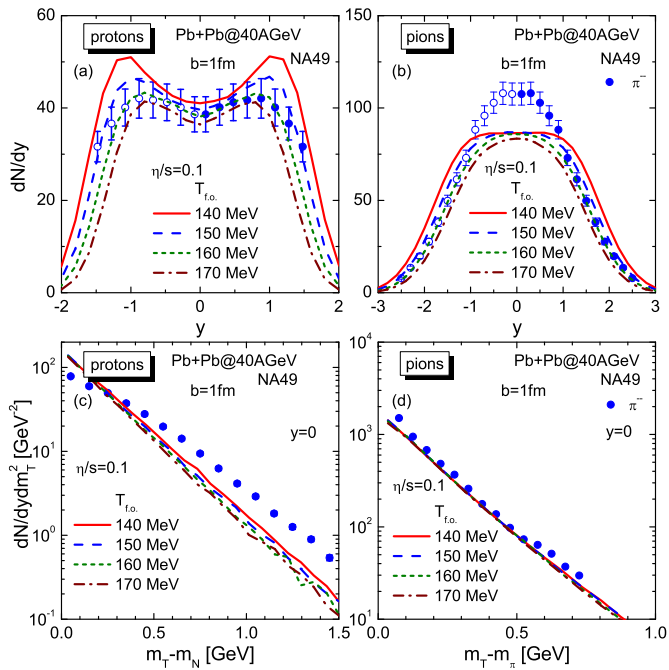}
\caption{
Rapidity distributions and transverse momentum spectra for protons and pions produced in Pb+Pb collisions at 40\,$\agev$ in comparison with the calculations of the HydHSD model for various values of the freeze-out temeperature $\Tfrz$. The viscosity is fixed at $\eta/s=0.1$. Experimental points are taken from Refs.~\cite{SPSN1,SPSN2,SPSpiK}.
} \label{fig:40AGeV-Tfo}
\end{figure}

The influence of the freeze-out temperature, $\Tfrz$, on rapidity distributions and transverse momentum spectra at mid-rapidity is illustrated in Fig.~\ref{fig:40AGeV-Tfo}. As is seen in the figure, proton
rapidity distributions become higher if the freeze-out temperature $\Tfrz$ is lower. The value of  $dN/dy$ at $y=0$  is moderately sensitive to $\Tfrz$ as well as to $\eta/s$, as we discussed above. This value is mainly determined by the hydrodynamics start time, $t_{\rm start}$.
Confronting Fig.~\ref{fig:40AGeV-Tfo}a and Fig.~\ref{fig:40AGeV-eta}a, we observe a similarity in the effects caused by a decrease of $\Tfrz$ and an increase of $\eta/s$, both lead to a growth of the humps in the proton $y$-distribution. The reason of that is an increase of the evolution duration in both cases  and a correlation of the parameters. A larger viscosity leads to a slower drop of the cell temperatures which increases the number of frozen cells in similar way as a choice of a smaller freeze-out temperature, see Fig. 11 below and the corresponding discussion. In the viscous case the two-hump structure of proton rapidity distribution at $\Elab=40\,\agev$ is clearly seen for any $\Tfrz$ (compare with Fig.~4b from \cite{HYDHSD2015}).

The width of the pion rapidity distribution is larger for smaller values of $\Tfrz$, as is seen in Fig.~\ref{fig:40AGeV-Tfo}b, whereas the distribution height is weakly dependent on $\Tfrz$. Also we observe saturation of  the height of the pion $y$-distribution with decreasing $\Tfrz$ similar to the dependence on $\eta/s$. As can be seen in Figs.~\ref{fig:40AGeV-eta}b, \ref{fig:40AGeV-Tfo}a, and \ref{fig:40AGeV-Tfo}b,  the height of the pion rapidity distribution saturates at the level which is significantly below the experimental data at mid-rapidity.
As a result, our calculations for $\Elab=40\agev$ can reproduce only the proton rapidity distribution but not the pion one if we vary both $\eta/s$ and $\Tfrz$ parameters.

Irrespectively to the saturation of  the pion distribution height with a decrease of $\Tfrz$, Fig.~\ref{fig:40AGeV-Tfo} demonstrates that there is an internal tension in attempts to describe simultaneously the proton and pion rapidity distributions in our model.
Let us show how it can be explained, basing on results for $\Elab=40\agev$. The reason of this failure in reproducing both distributions is the discussed-above insensitivity of the $y$-distributions for $\eta/s>0.1$. Therefore, after the increase in the distribution by the variation of $\eta/s$ is exhausted, we have only one parameter $\Tfrz$ to tune both proton and pion distributions.  So for a freeze-out temperature, $140\,{\rm MeV}\lsim \Tfrz \lsim 160$\,MeV, which is needed to fit the proton rapidity distribution, we have only the correct width of the pion distribution.

Transverse momentum spectra of protons and pions at mid-rapidity ($y=0$) are shown in Figs.~\ref{fig:40AGeV-Tfo}c and~\ref{fig:40AGeV-Tfo}d, respectively, for various values of freeze-out temperature. The striking feature is that the slope of the pion spectra is almost insensitive to the variation of $\Tfrz$ and the proton spectra demonstrate very weak dependence, whereby the slope's steepness decreases with a $\Tfrz$. As the result, to approach experimental data for the proton spectrum we have to choose $\Tfrz<120$\,MeV, whereas $\Tfrz\simeq 150$ MeV is necessary for the description of the proton rapidity distribution.
In contrast, the pion $m_{\rm T}$ spectrum is well described by our model.

\subsection{Constraints on the shear stress tensor}
\label{VHLLEMUSIC}

The above results lead to two questions. Why viscous effects in our 2-stage hybrid model for pion rapidity distribution are so small ($\sim$10\%) while the results of the authors \cite{KHPB} within the vHLLE+UrQMD model demonstrate that the response is large (about 20\%, see Fig.~4 in the cited article)? It cannot be explained by taking into account the electric charge conservation since this effect is included in both ideal and viscous versions of the model~\cite{KHPB}. The second question is why our model is insensitive to the $\eta/s$ value  at $\Elab=40\agev$, if $\eta/s>0.1$, see Fig.~\ref{fig:40AGeV-eta} above.

\begin{figure}
\centering
\includegraphics[width=8.8cm]{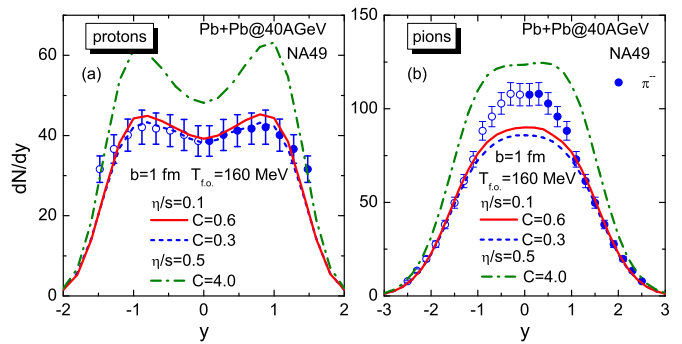}
\caption{
Proton (a) and pion (b) rapidity distributions at $\Tfrz=160$~MeV and various values of the $C$ parameter in the shear-stress tensor constraint (\ref{piconstrain}): $C=0.3$ and 0.6 for $\eta/s=0.1$, and $C=4$ for $\eta/s=0.5$.
The experimental data are the same as in Figs.~\ref{fig:40AGeV-eta} and~\ref{fig:40AGeV-Tfo}.
} \label{fig:40AGeV-C}
\end{figure}

First of all, we check how the viscous response is changed if the constant $C$ in Eq. (\ref{piconstrain}) is increased. When $C$ is larger, the viscous effects are expected to be more pronounced. As Fig.~\ref{fig:40AGeV-C} shows, this is indeed the case. From Fig.~\ref{fig:40AGeV-C}a we see that a two-hump structure in proton rapidity distributions is more pronounced for larger values of $C$-parameter, while Fig.~\ref{fig:40AGeV-C}b demonstrates that simultaneously pion rapidity distribution is getting a bit higher but the gain at mid-rapidity is too small to improve noticeably the agreement with the experiment.

To study the behaviour of the code at large $\eta/s$ we relax the constraint  (\ref{piconstrain}) choosing the $C=4$ and $\eta/s=0.5$. The humps in proton $y$-distribution is growing up further and the pion distribution increases significantly. The reason for this strong increase will be discussed later in this section. In further we will not consider explicitly calculations with $C>1$ as we believe that solving hydrodynamic equation beyond their applicability range makes no scientific sense. Viscous corrections at particle freeze-out, which can become anomalously large, and nonphysical reheating of fluid elements are just several problems to be named.  Moreover, our calculations show that the total energy of \underline{produced} particles becomes at the end larger than the initial total energy of the system if an unlimited increase of $C$ parameter is allowed. The reason for this is that due to reheating, some cells, which were frozen-out at some intermediate step, gain energy and become 'active', and then freeze-out again on the later stage.  Such multiple inclusions of recycled cells to the freeze-out hypersurface is supported by an extension of the system evolution at larger values of $C$. Another reason, which can play a more important role, is that due to reheating, cells intersect freeze-out hypersurface later, when the volume of the fireball is larger. As the result the volume hypersurface is higher than for the case without reheating, see also discussion of Fig.~\ref{fig:centcell}.
{\it Thus, one can suggest that the tracking of the total energy of produced particles could be used to put a physical restriction on the maximal value of $C$ for the given collision energy and the code.}  Many authors, see \cite{MNR2010,KHB2013,DuHeinz19} suppose that viscous corrections have to be small, i.e. $C\leq 1$. We also follow this trend and use $C<1$ for our standard set of parameters.

Second, we have to note that the vHLLE model~\cite{KHB2013} uses another constraint on the $\pi^{\mu\nu}$ tensor magnitude using the criterion (\ref{piconstrain}) where the quantity $q$ is calculated as
\begin{align}
\label{pivHLLE}
q=q_{\rm V}\equiv \frac{\max_{\mu,\nu} \left|\pi^{\mu\nu}\right|}{\max_{\mu,\nu}\left|T_{\rmid}^{\mu\nu}\right|}\,.
\quad \mbox{(V-cond.)}
\end{align}
It results in a weaker condition than with our definition~(\ref{q-def-S}).
We will denote the condition used in the vHLLE model as the V-condition.
For a further comparison we also consider the condition  which is applied in the MUSIC model~\cite{MUSIC}, where one defines
\begin{align}
\label{pimusic}
q=q_{\rm M}\equiv \sqrt{\frac{\pi^{\mu\nu}\pi_{\mu\nu}}{T_{\rm id}^{\mu\nu}T_{\rm id,\mu\nu}}}
\,. \quad \mbox{(M-cond.)}
\end{align}
We will call it the M-condition\footnote{As one can see from the function {\it QuestRevert} of MUSIC code, the developers use an energy-dependent cut-off parameter $C=C(\varepsilon)$ in Eq.~(\ref{piconstrain}). We take just a constant value.}.

\begin{figure}
\centering
\includegraphics[width=8.8cm]{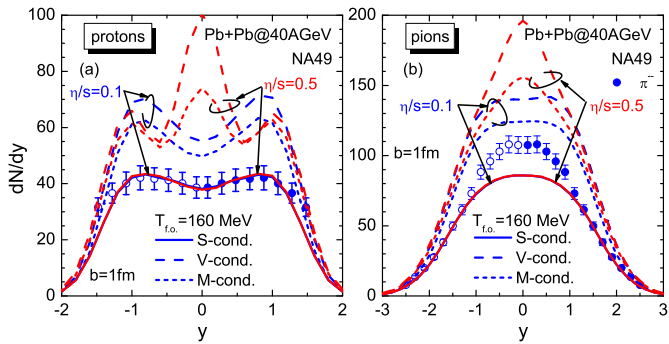}
\caption{Rapidity distributions of protons (a) and pions (b) for Pb+Pb collisions at $\Elab=40\,\agev$ for different choices of $\pi^{\mu\nu}$ constraints (\ref{q-def-S}), (\ref{pivHLLE}), and (\ref{pimusic}). Calculations are carried out for the freeze-out temperatures $\Tfrz=160$\,MeV and two values of the shear viscosity $\eta/s=0.1$ and $0.5$. The experimental data are the same as in Figs.~\ref{fig:40AGeV-eta}-\ref{fig:40AGeV-C}.} \label{fig:40AGeV-pi}
\end{figure}

Both V- and M- conditions can be easily realized in our code.
The results of calculations for Pb+Pb collisions at $\Elab=40\,\agev$ with $\Tfrz=160$~MeV, $\eta/s=0.1$ and 0.5, and $C=0.3$ are shown in panels~(a) and (b) of Fig.~\ref{fig:40AGeV-pi} for protons and pions, respectively. One can see that applying weaker constraints, Eqs.~(\ref{pivHLLE}) or~(\ref{pimusic}), lead to a dramatic change in the rapidity distributions. For $\eta/s=0.1$ the height of the proton humps and the maximum of the pion distribution increase sizably compared to the calculations with the stricter constraint (\ref{q-def-S}). Also, Fig.~\ref{fig:40AGeV-pi} demonstrates that the sensitivity of the rapidity spectra to the $\eta/s$ value is much larger for the V- and M-conditions than for the S-condition.  This property is kept even if we increase the constant $C$ in the S-condition until $C=4.0$. So V- and M-constraints also more sensitive than S-one. There appears even a three-hump structure in the proton rapidity distribution for $\eta/s=0.5$ when one applies the V- or M-conditions. For the S-constraint such a structure does not appears even if one let the code run with $C=4$ and $\eta/s=0.5$.
We see in Fig.~\ref{fig:40AGeV-pi}b that with the weaker constraints one can reproduce the mid-rapidity dip in the pion rapidity distribution by changing the $\eta/s$ parameter.

The effect of the $\pi^{\mu\nu}$ constraint relaxing  in comparison to the the strict condition (\ref{piconstrain}) and (\ref{q-def-S}) is qualitatively similar for the V- and M-conditions; however for the V-condition the effect is more pronounced and increases strongly for a larger value of $\eta/s$. This property of two weaker conditions is found to be valid for all considered energies.

\begin{figure}
\centering
\includegraphics[width=7cm]{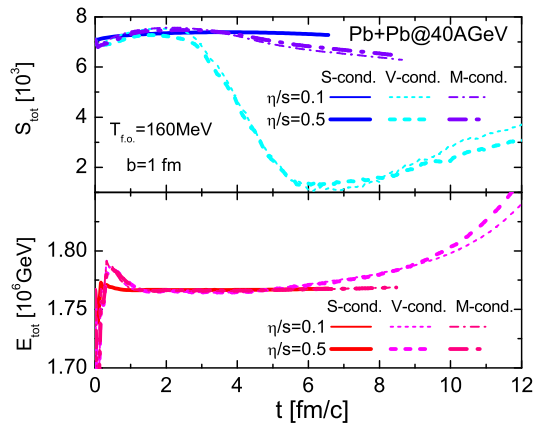}
\caption{Evolutions of the total entropy  and total energy for different choices of $\pi^{\mu\nu}$ constraints for Pb+Pb collisions at $\Elab =40\,\agev$ with $\Tfrz=160$\,MeV for $\eta/s=0.1$ and $0.5$.}
\label{fig:EStot-pi}
\end{figure}

Figure~\ref{fig:EStot-pi} demonstrates that different constraints on the $\pi^{\mu\nu}$ tensor affect also the evolution of such global quantities as the total energy and the total entropy of the system. A weaker constraint leads to longer system evolution  time, especially for V-condition. For the S- and M-constraints the total energy stays constant with good precision, whereas for the V-condition the total energy starts to increase after the first 5\,fm/$c$ and continues its growth reaching a $\sim 2\%$ excess at time 12\,fm/$c$. At the same time, the total entropy of the system decreases only slightly for the S- and M-conditions, but demonstrates a violent behaviour -- first a strong decrease, then a moderate  increase -- for the V-condition. Such a sharp entropy change for the V-condition occurs because of a large increase of the number of cells where the viscous correction to the entropy flow exceed the ideal part.

One should keep in mind that the entropy density is the ternary quantity in our case, where the primaries are the energy-momentum tensor components, the secondaries are the energy density, the baryon density, and the velocity. Therefore, the errors of the entropy calculations are quite large. 

If we let the code run with the S-condition $C=4$ and $\eta/s=0.5$ allowing thereby the untamed increase of the viscous term we find that the energy nonconservation becomes evident at earlier times ($\sim 4$\,fm/$c$) and the entropy starts to vary rapidly also. Therefore, the pattern seen in Fig.~\ref{fig:EStot-pi} for the V- and M-conditions can be unequivocally attributed to the uncontrolled promotion of viscous effects.

\begin{figure}\centering
\includegraphics[width=8.8cm]{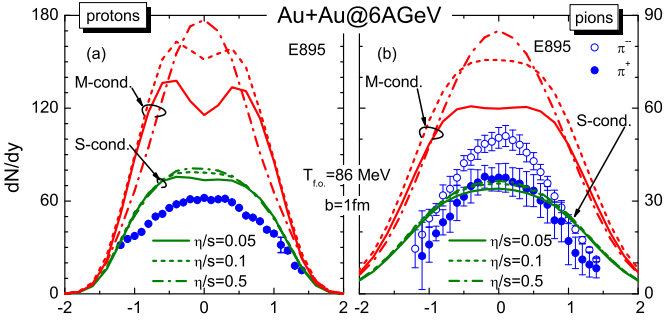}
\caption{ Rapidity distributions of protons and pions for $\Elab=6\,\agev$ with $\Tfrz=86$\,MeV at three values of $\eta/s=0.05$, 0.1, and $0.5$ and different conditions on the $\pi^{\mu\nu}$ tensor: the S-condition (\ref{q-def-S}) and the M-condition (\ref{pimusic}).
Experimental points are from~\cite{E895-prot,E895-pion}.}
\label{fig:6AGev-pi}
\end{figure}

To better understand how the form of the constraint on the shear stress tensor affects observables, let us consider also collisions at AGS energies.
The results for Au+Au collisions at $\Elab=6\,\agev$ are shown in Fig.~\ref{fig:6AGev-pi}, where we put $\Tfrz=86$\,MeV and consider three different viscosities with $\eta/s=0.05,\ 0.1$, and $0.5$. For simplicity, we do not take into account nucleon coalescence whose effect is quite small. As one can see, even for $\eta/s=0.05$ S- and M-conditions give different results for both pion and proton distributions. The distributions for the M-condition are higher and the two-hump structure appears in the proton distribution.
The difference is enhanced for lager values of $\eta/s$, and the two-hump structure in the proton $y$-distribution disappears at
$\eta/s=0.5$. Thus, the calculations for $\Elab=6\,\agev$ confirm our conclusions that weak $\pi^{\mu\nu}$ constraints lead to large sensitivity of a hydrodynamic model to the $\eta/s$ value.

\begin{figure}
\centering
\includegraphics[width=8.8cm]{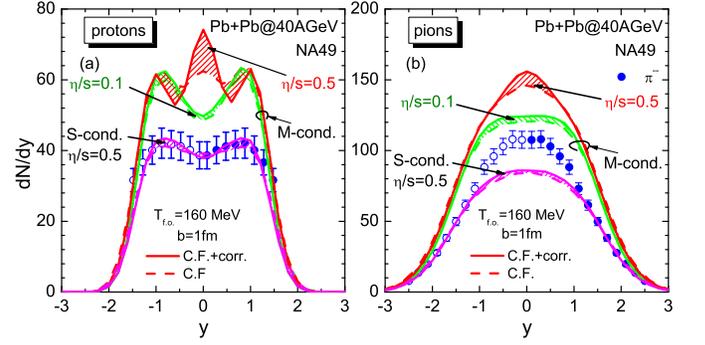}
\caption{Proton and pion rapidity distributions calculated using the Cooper-Frye formula
(\ref{CooperFrye}) with and without viscous corrections (\ref{viscdistr}) shown by solid and dashed lines, respectively. Calculations are performed for Pb+Pb collisions at $E_{\rm lab}=40\,\agev$ with the M-condition and for $\eta/s=0.1$ and 0.5 and for the S-condition with $\eta/s=0.5$. The results for $\eta/s=0.1$ would be very similar if the S-condition is applied.
The experimental data are the same as in Figs.~\ref{fig:40AGeV-eta}-\ref{fig:40AGeV-C}. }
\label{fig:CF-corr}
\end{figure}

\begin{figure}
\centering
\includegraphics[width=6cm]{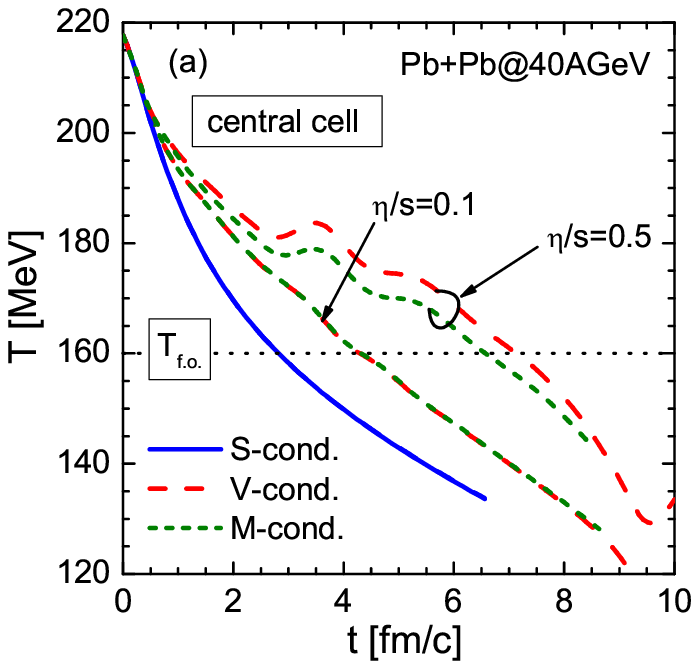}
\includegraphics[width=6cm]{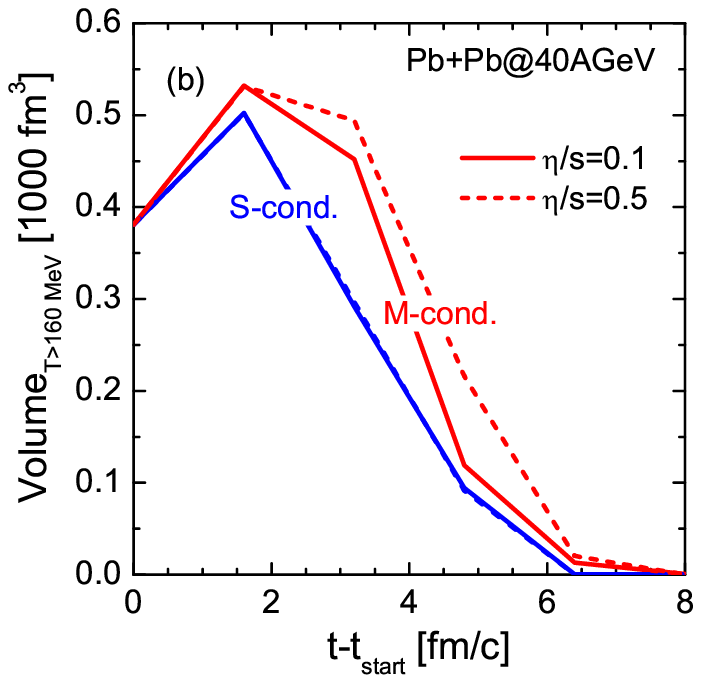}
\caption{(a)~Evolution of the temperature in the central cell.
(b) The evolution of the volume of cells with temperatures $T>T_{\rm f.o.}=160$\,MeV. Calculations are performed for Pb+Pb collision at $E_{\rm lab}=40\,\agev$ and
for two values of $\eta/s$ and various conditions on the shear stress tensor.
}
\label{fig:centcell}
\end{figure}

The obtained results confirm our earlier conclusion in Ref.~\cite{HYDHSD2015} that a two-hump structure in proton and pion distributions has the kinematic (dynamic) origin and is not necessarily related to a phase transition.

In Fig.~\ref{fig:CF-corr} we illustrate the role of the viscous correction term in the Cooper-Frye formula, see Eq.~(\ref{viscdistr}), where we present the proton and pion rapidity distributions calculated with and without the last term in square brackets in (\ref{viscdistr}). For pions this corrections are truly perturbative leading to a slight increase of the pion number at mid-rapidity. For calculations with the M-condition the effect is stronger than for those with the S-conditions and increases with the $\eta/s$ growth.
For protons, however, in calculations with $\eta/s=0.5$ the correction terms in the Cooper-Frye lead to a change in the shape of the distribution making the three-hump structure more pronounced.

Why the viscous effects promoted by the weak constraint with the V- and M-conditions lead to an increase in the pion number multiplicity? To answer this question we show in Fig.~\ref{fig:centcell}a the evolution of the central cell temperature for calculations done with various conditions. The viscous effects prolongs the evolution and increase the temperature. Even some reheating effect for the central cells are seen for runs with V- and M-conditions. Also the number of fluid cells with temperatures $T>T_{\rm f.o.}=160$\,MeV increases for runs with weaker conditions, as illustrated in Fig.~\ref{fig:centcell}b for the M-condition. An increase of the specific viscosity results in a further increase of the volume. The combination of higher temperatures and larger freeze-out volume leads to strong increase of the number of pions (not restricted by any conservation law) if the viscous effects are constrained by V- and M-conditions.

\begin{figure}[h]
\centering
\includegraphics[width=6cm]{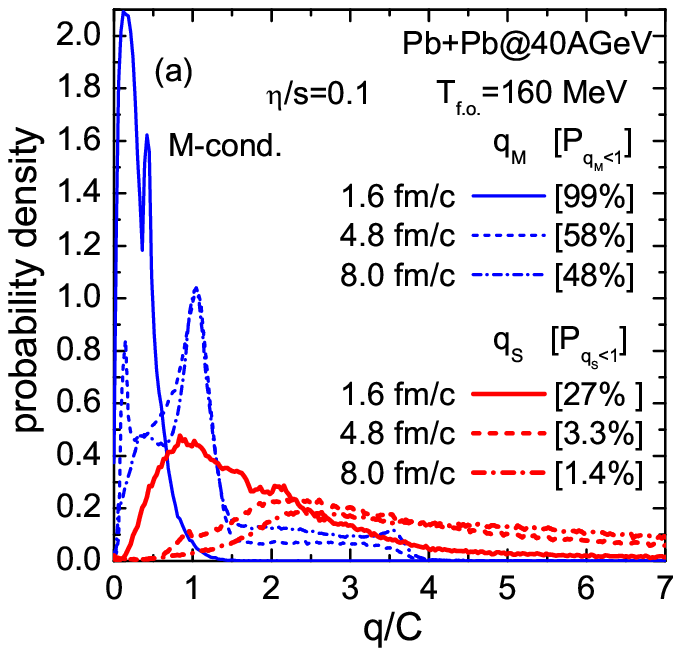}\\
\includegraphics[width=6cm]{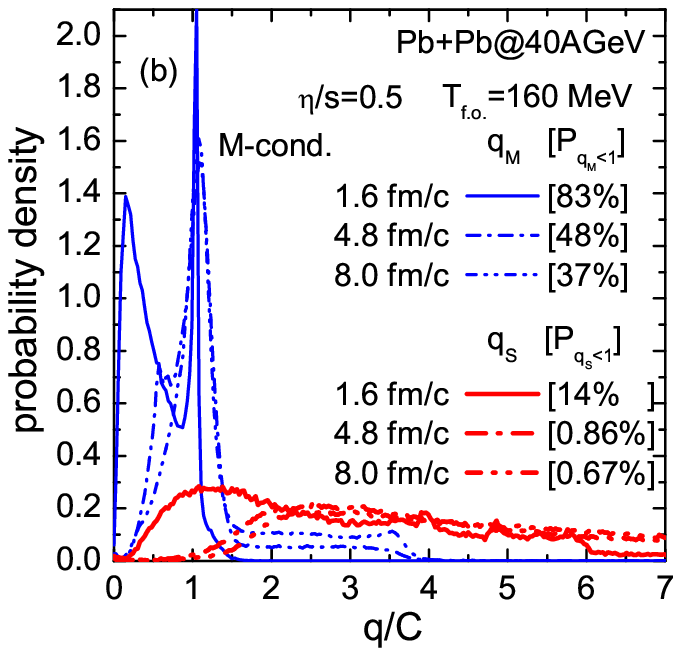}
\caption{Probability density to find a fluid cell in the system with particular values of $q_{\rm M}$ (thin lines) and $q_{\rm S}$ (thick lines) parameters defined by Eqs.~(\ref{pimusic}) and~(\ref{q-def-S}) in hydrodynamic runs for the Pb+Pb collisions at $E_{\rm lab}=40\,\agev$ performed with the M-condition. Panel (a) shows the result for $\eta/s=0.1$ and panel (b) for $\eta/s=0.5$. Lines are shown for three values of $t-t_{\rm star}=1.6$, 4.8, 8.0\,fm/$c$. Numbers in the square brackets show probability to find a cell with $q_{\rm M(S)}/C<1$.
}
\label{fig:q-distrib}
\end{figure}
It is interesting to try to quantify to what extend the viscous effects remains perturbative in the course of hydrodynamic evolution. With this aim we run the code for Pb+Pb collision at $E_{\rm lab}=40\,\agev$ and calculated the distribution of the values of $q_{\rm M}$ among all fluid cells with temperatures $T>100$\,MeV. The normalized (in the interval $0\le q/C\le 10$) distributions obtained for three different moments of time are shown by thin lines in Fig.~\ref{fig:q-distrib} for runs with $\eta/s=0.1$ and $\eta/s=0.5$. We see that initially the majority of cells, $\sim 99\%$ for $\eta/s=0.1$ and $\sim 83\%$ for $\eta/s=0.5$, have $q_{\rm M}<1$ but already in the middle of the fireball evolution $t-t_{\rm start}=4.8\,{\rm fm}/c$ the maximum of the distribution is at $q_{\rm M}\sim 1$ and the rescaling (\ref{rescale}) of the $\pi^{\mu\nu}$ tensor must be performed in the range of about 40 to 50\% of cells. At the final stage, however, more then 60\% of cells are rescaled in the case of $\eta/s=0.5$. This results are obtained for the codes running with the M-condition. The true characteristic for perturbativity of the viscous effect is however the quantity $q_{\rm S}$. If $q_{\rm S}>1$ then at least one of the elements in the $\pi^{\mu\nu}$ tensor is larger than the corresponding element in $T_{\rm id}^{\mu\nu}$, i.e. the viscous effect is non-perturbative and the applicability of the hydrodynamic equations (\ref{hydrobase}) and (\ref{ISeqs}) is questionable. The distributions of values $q_{\rm S}$ are shown in Fig.~\ref{fig:q-distrib} by thin lines. We see that although the code is keeping $q_{\rm M}<1$ at each evolution step, vast majority of fluid cells have $q_{\rm S}>1$. So, already at initial steps only in 27\% for $\eta/s=0.1$ and in 14\% for $\eta/s=0.5$  of all cells the viscous effects are truly perturbative. With time passed these numbers drop further down to a very small values: at $t-t_{\rm start}=8\,{\rm fm}/c$ they are 1.4 and 0.67\% for $\eta/s=0.1$ and 0.5, respectively.
So, applying the weak M-condition we let the hydrodynamic code run, in reality, in the non-perturbative regime already at $\eta/s = 0.1$.
We find that in such calculations the total energy of produced particles, $E_{\rm prod}$, exceeds the initial total energy of the fireball, $E_{\rm tot}(t_0)$. (One should note that a weakening of constraint on viscous term similar to the M(V)-condition can be also obtained for the S-condition if one lets $C$ be quite large.) This indicates that one cannot use the Israel-Stewart equations (\ref{ISeqs}), and one needs to include higher-order gradient terms on the r.h.s. This extension we will consider separately elsewhere. Also in such a regime it might be necessary to take into account additional terms in Eq.~(\ref{viscdistr}).
For the S-condition, the situation is different. On one hand,
we observe that rapidity distribution becomes insensitive to an increase of $\eta/s$ for $\eta/s>0.1$. On the other hand the energy of produced particles, $E_{\rm prod}$ is always less then the total initial energy $E_{\rm tot}(t_0)$ even at $\eta/s=0.5$. Therefore, we tend to rely more on the results obtained with the S-condition, whereas the result obtained with the M-condition are not trustworthy for $\eta/s> 0.1$. However, as we will see in the next Section, only the fit with the M-condition at $\Elab=10.7\ \agev$ demands $\eta/s$ essentially higher than $0.1$.

\section{Beam-energy dependence of parameter influence}
\label{fitsection}

\begin{figure*}
\centering
\includegraphics[width=17cm]{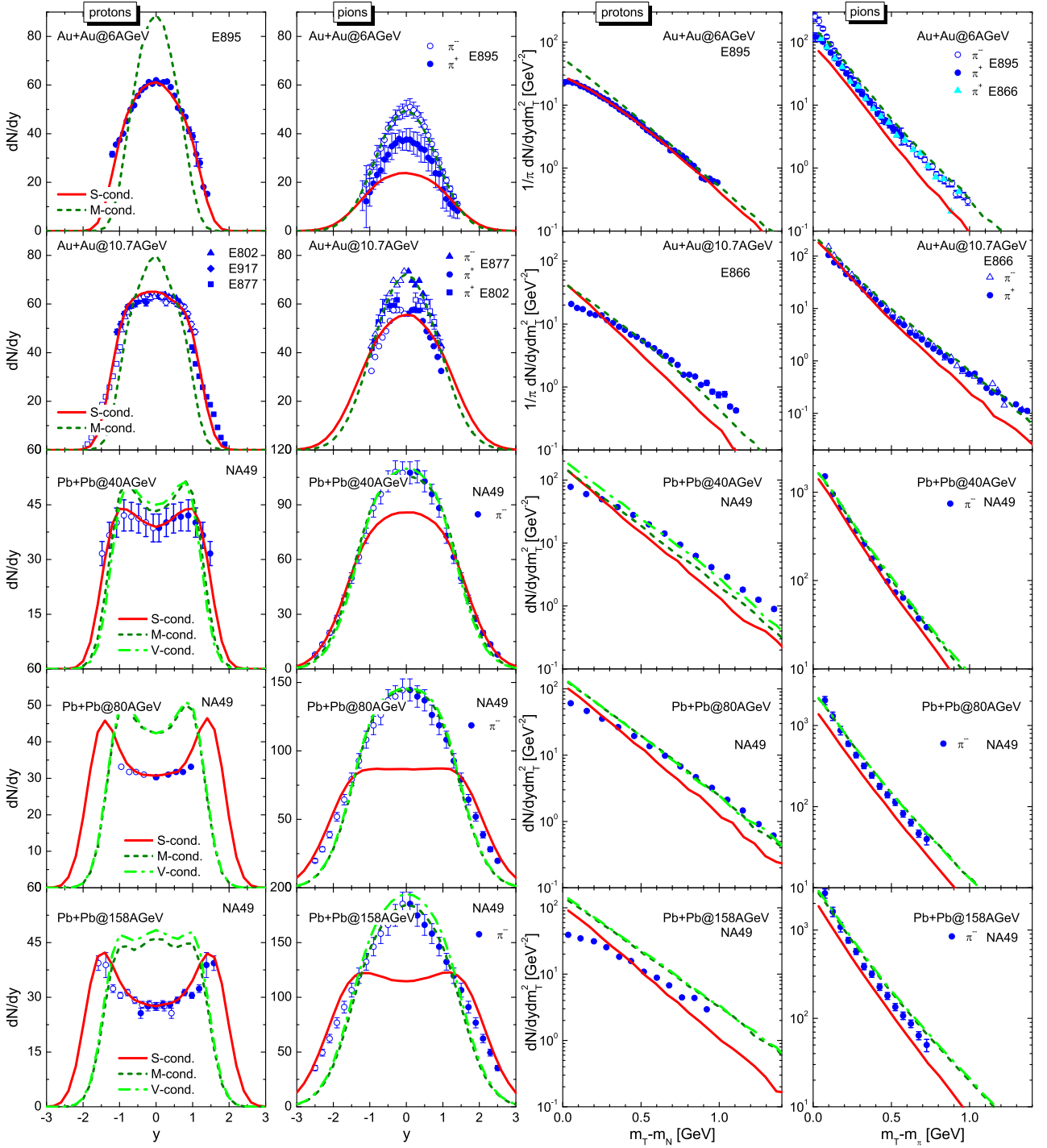}
\caption{Best fits of rapidity distributions and mid-rapidity transverse momentum spectra of protons and pions obtained by varying the values of parameters $\eta/s$ and $\Tfrz$. The best values of fitting parameters are given in Table~\ref{tab:fit-parameters}. Experimental data are from Refs.~\cite{
SPSN1,SPSN2,SPSpiK,E895-prot,E895-pion,PRC57,PRC62,PRL86,PRC59,NPA610, PRC73, JPhysG35, NPA715}.}
\label{fig:best-fits}
\end{figure*}

After considering the properties of different $\pi^{\mu\nu}$ constraints, we can try to fit the rapidity distributions and transverse momentum spectra in a wide range of bombarding energies reachable at the AGS and SPS facilities. The results of this fit are presented in Fig.~\ref{fig:best-fits}. For each energy we vary independently the parameters $\Tfrz$ and $\eta/s$. We also try various conditions on the shear stress tensors evaluating the quantity $q$ in the constraint~(\ref{piconstrain}), keeping there $C=0.3$, according to the strict S-condition~(\ref{q-def-S}) and weaker V- (\ref{pivHLLE}) and M-conditions (\ref{pimusic}).
The obtained best values of the varied parameters are collected in Table~\ref{tab:fit-parameters}.

\begin{table*}[!ht]
\caption{The fitted parameters for Eq. (\ref{piconstrain}) (S-condition) and for Eq. (\ref{pivHLLE}) if proton [(p)V-condition] or pion [($\pi$)M-condition and ($\pi$)M-condition] rapidity distribution is tuned (The fit accuracy is $\Delta T=\pm 5-10 MeV$, $\Delta \eta/s=\pm0.05$).}
\label{tab:fit-parameters}
\begin{tabular}{cccccccccccc}
\hline\hline
& \multicolumn{2}{c}{S-condition} && \multicolumn{2}{c}{$(\pi)$M-condition} && \multicolumn{2}{c}{$(\pi)$V-condition} && \multicolumn{2}{c}{(p)M-condition} \\
\cline{2-12}
\raisebox{1.6ex}{$\Elab\,[\agev]$}
&$\Tfrz$\,[MeV] & $\eta/s$  && $\Tfrz$\,[MeV] & $\eta/s$ && $\Tfrz$\,[MeV] & $\eta/s$ && $\Tfrz$\,[MeV] & $\eta/s$\\
\hline\hline
6    & 78  & 0.03  && 120  & 0.07  && - & - &&  79  & 0.01 \\
\hline
10.7 & 125 & 0.01 && 143 & 0.24 && - & - &&  125  & 0.01 \\
\hline
40   & 155 & 0.05 && 170 & 0.08 && 175 & 0.1 &&  160  & 0.01 \\
\hline
80   & 160  & 0 && 190 & 0.13 && 190 & 0.11 &&  160  & 0  \\
\hline
158  & 160 & 0    && 205 & 0.1    && 205 & 0.08 &&  160  & 0 \\
\hline\hline
\end{tabular}
\end{table*}

Consider, first, the results obtained with the S-condition when we fit proton rapidity distributions shown in Fig.~\ref{fig:best-fits} by solid lines. As we have already seen in the previous sections, we cannot simultaneously reproduce pion and proton distributions  in this case.
For collisions with energies from  6\,$\agev$ to  158\,$\agev$,  we can properly reproduce proton rapidity distributions. In agreement with our previous results for ideal hydrodynamics \cite{HYDHSD2015}, where we also well reproduced proton rapidity distribution, we obtain small values of $\eta/s\simeq 0-0.05$. At all considered energies the experimental pion rapidity spectra are underestimated for $-1\lsim y\lsim 1$.  Different values of $\Tfrz$ are found for different energies. One can see that obtained $\Tfrz$ values are quite close to predicted by a thermal statistical model~\cite{Andronic} and demonstrate a saturation at higher energies.
Once we have managed to fit the proton $y$-distributions, the proton $m_T$-spectra have typically too steep slopes, except the $E_{\rm lab}=6\,\agev$ case, where both $y$- and $m_T$-distributions can be reproduced reasonably. Although, even in this case the number of pions at midrapidity is too low. The
slopes of the pion $m_{\rm T}$ spectra are typically closer to the experiment than those for proton ones, except the 10.7\,$\agev$ case, but the lines systematically go below the data.

We would like to note that evidence for similar small values of $\eta/s\sim0.04$ were also found in Ref.~\cite{smalletas} at RHIC energies from an analysis of the elliptic flow of charged hadrons.

Now we try to apply the weaker M-condition (\ref{pimusic}). Here we can follow two strategies: we can insist on fitting at best either proton $y$-distributions or the pion ones. We denote the results obtained in the first way  as (p)M-fits and as ($\pi$)M-fits in the second case.
It turns out that the results of (p)M-fits are close to those obtained with the S-conditions. The fit parameters are also similar, see last 2 columns in Table~\ref{tab:fit-parameters}.

The situation changes if we require the best possible description of pion rapidity distributions.  This is possible with the M-condition since, as demonstrated in Section~\ref{VHLLEMUSIC}, the code in this case does not lose its sensitivity to the viscosity parameter, at cost of the uncontrollable increase of the number of cells where the elements of the $\pi^{\mu\nu}$ tensor exceed dramatically the components of the $T_{\rm id}^{\mu\nu}$ tensor.  As is seen from  Table II, in this case we obtain systematically higher freeze-out temperatures and viscosities than for the S-constraint. Obtained $\Tfrz$ lies very far from those predicted by statistical model~\cite{Andronic}. Also for the M-condition we found that the parameter $\eta/s$ has a maximum at $E_{\rm lab}=10.7\agev$ while for the fit of proton distributions with the S-restriction, we maybe observe a minimum (due to an inaccuracy in the fitting procedure we can have a constant). Applying the M-condition,
as one can see from Fig.~\ref{fig:best-fits},  we are able nicely reproduce both $y$- and $m_{\rm T}$-distributions of pions.
However, if we look at the proton rapidity distribution (not shown in~\cite{KHPB}) we see that we miserably fail:  the number of protons near the mid-rapidity is too high and the distributions are too narrow (for all energies excluding 80$\agev$). These are the signals of the strong viscosity effects, as we discussed in Section~\ref{VHLLEMUSIC}, see Figs.~\ref{fig:40AGeV-pi} and~\ref{fig:6AGev-pi}.

The same results we are also obtained if we use the V-condition~(\ref{pivHLLE}) instead of the M-condition, see dash-dotted lines in Fig.~\ref{fig:best-fits}. As we see both weak conditions give very close results if one fits pion rapidity distributions. It can be easily explained since V-condition can be presented in the equivalent quadratic form as sum of squared components of $\pi^{\mu\nu}$ on l.h.s and $T^{\mu\nu}_{\rm id}$ on the r.h.s. The resulting expressions differs from the M-condition only by some ``$+$'' signs instead ``$-$'' in the latter.

Summarising the discussion of Fig.~\ref{fig:best-fits}, we conclude that any considered condition does not allow to reproduce simultaneously pion and proton experimental data.

In the discussed AGS-SPS energy range, a detailed comparison of experimental data with different viscous hydro approaches has been made only in a couple of papers. Great success was reached in terms of the three-fluid dynamics (3FD) model~\cite{IRT06} applied to energies $E_{\rm lab}\lsim 158\,\agev$. The 3FD approximation is a minimal way to simulate the early-stage nonequilibrium in colliding nuclei. In contrast to the conventional 1-fluid hydrodynamics, the 3FD approach takes into account a finite stopping power in a counterstreaming
regime of leading baryon-rich matter at an early stage of a collision, which allows one to use a constant  $t_{\rm start}$ parameter independently of  $\sqrt{s}$. Different EOS are used instead of parameter variation being in the best agreement for the case of smooth cross-over phase transition. The beam-energy dependence of rapidity (not pseudorapidity !) proton spectra is in  good agreement with experiment~\cite{Iv16} at $E_{\rm lab}\lsim 10\,\agev$, for all EOS, but a mixed phase with the smooth crossover dominates definitely at higher energies. A similar situation is with the transverse mass spectra at the middle rapidity~\cite{Iv14}. Effects of the EOS are getting visible in more delicate characteristics,  say, energy dependence of the slopes of transverse mass spectra for identified hadrons.

The collective behavior of the nuclear fireball can also be studied using the hydrodynamics inspired phenomenological model called the blast wave model \cite{FB04}. The main underlying assumption of this model is that the particles in the system produced in the collisions are locally thermalized and the system expands collectively with a common radial velocity field undergoing an
instantaneous common freeze-out. While the spherically expanding source may be expected to mimic the fireball created at low energies, at higher energies a stronger longitudinal flow might lead to cylindrical geometry. For the latter case, an appropriate formalism was first developed in Ref.~\cite{SSH93}. Using a simple functional form for the phase space density at kinetic freeze out, the authors approximated the hydrodynamical results with the boost-invariant longitudinal flow.  The common assumption for all variants of the blast
wave model is the underlying boost-invariant longitudinal dynamics. Although it is a reasonable assumption at RHIC and LHC energies, longitudinal boost-invariance does not hold well at AGS-SPS energies. Recently, a non
boost-invariant  blast wave model has been developed~\cite{RBJR18}.
The model was successfully used in the AGS-SPS energy range to fit the rapidity distributions and transverse momentum spectra with only two parameters, namely, a kinetic freeze-out temperature $T_{\rm f.o.}$ and a radial flow strength $\beta_T$.
The fitted here $T_{\rm f.o.}$ are smaller than the values in our analysis (see Table II). One should note that the model~\cite{RBJR18} nicely describes the shape of distributions studied but their absolute values should be separately fitted at every energy.

\section{Conclusions}

In this work we have extended the HydHSD model developed in~\cite{HYDHSD2015}
by inclusion of shear viscosity within the Israel-Stewart hydrodynamics. Using the updated version of our hybrid model, we considered proton and pion rapidity distributions and transverse momentum spectra for $E_{\rm lab}\leq 160\agev$.
As in other viscous hydrodynamic calculations, genuine inaccuracy of a numerical implementation leads to an uncontrollable increase of the shear stress tensor $\pi^{\mu\nu}$ that contradict to a perturbative character of the viscous corrections to ideal hydrodynamics.
An observable consequence of this is that the energy of produced (frozen-out) particles can increase the total initial energy making thereby the results of such calculations unreliable. Also, without such a control codes run in numerical instabilities~\cite{Denicol18}.

To timid the problem, a regularization scheme was suggested in the literature, which assumes a rescaling of the $\pi^{\mu\nu}$ if some condition of perturbativity of $\pi^{\mu\nu}$ in comparison to the ideal energy-stress tensor $T^{\mu\nu}_{\rm id}$, cf. Eq~(\ref{T-ideal}), is violated.
We use the strict (S-) condition (\ref{q-def-S}) proposed in \cite {MNR2010}, which guarantees  that each element of the $\pi^{\mu\nu}$ tensor remains not more than 30\% (speciefed by the $C$ value in (\ref{piconstrain})) of the corresponding element of the $T_{\rm id}^{\mu\nu}$ tensor. Also we analyzed other conditions used in the literatures:
the V-condition (\ref{pivHLLE}) used in the vHLLE code~\cite{KHB2013,KHPB}
and the M-condition (\ref{pimusic}) used in the MUSIC and  iEBE-VISHNU codes~\cite{MUSIC,VISHNU}. The V- and M-conditions are weaker then
S-condition. This weakness is confirmed also numerically since the similar results for rapidity distributions can be obtained with the S-condition for a quite large value of $C$.

It is proven that the form of $\pi^{\mu\nu}$-constraints plays a crucial role in sensitivity to the $\eta/s$ value. If we fix a form of the constraint, the $\eta/s$-sensitivity increases for larger $C$. Our results demonstrate that the form of the proton humps is mainly determined by the $\eta/s$ ratio which has to be not greater then $0.1$ for $\Elab=40\agev$, see Fig.~\ref{fig:40AGeV-eta}.

A numerical algorithm with weaker V- and M-conditions is more responsive and leads to higher pion rapidity and humps in proton distribution than with the stricter S-condition for the same freeze-out parameters, see Figs.~\ref{fig:40AGeV-pi}, \ref{fig:6AGev-pi}, and~\ref{fig:CF-corr}. Our results demonstrate that the V-condition leads to the longest evolution and,
as a result, is most sensitive to a change of the $\eta/s$ parameter; however, they cannot be applied to large values of the $\eta/s$.

Such sensitivity of weaker conditions allows for a good quality fits of pion rapidity distributions. The reason for this is that the larger viscosity effects, going beyond perturbative nature of the original hydrodynamic equations, lead to higher temperatures of fluids and consequently to a higher freeze-out volume contributing to the pion yield, see Fig.~\ref{fig:centcell}. The price of the weakening is that the total energy of produced particles becomes greater than the initial energy already at $\eta/s\sim 0.1$. This effect can be used to determine the physical restriction on the value of the parameter $C$ for given collision energy and shear viscosity.

However, any considered condition does not allow to reproduce simultaneously pion and proton experimental data within our model. Moreover, a particular parameter fitting of experimental distributions at every colliding energy $E_{\rm lab}$ does not guarantee excellent agreement.
We expect that quantitative improvement of the description could be reached by choosing a better equation of state and by taking into account fluctuating (event-by-event) initial conditions and the bulk viscosity.
Note that for the considered moderate beam-energy range there is no systematic comparison of hydro-predictions with experimental data though for separate observable good agreement with experiment may be reached.

\vspace{3mm}
\begin{acknowledgments}
We thank E.~Bratkovskaya and W.~Cassing for providing the HSD code and consultations. We appreciate very much extensive discussions with  Iu.~Karpenko and Yu.B.~Ivanov and constructive remarks by G.~Sandukovskaya.
The work is supported by Slovak grant VEGA-1/0348/18 and by THOR the COST Action CA15213. A.S.K and E.E.K. acknowledge the support by the Plenipotentiary of the Slovak Government at JINR, Dubna. The work of A. Khvorostukhin was supported by the RFBR grant no. 18-02-40137.
\end{acknowledgments}

\appendix

\section{Reconstruction of local quantities}
The hydrodynamics code evolves the components of the energy-stress tensor and baryon current. The equation of state is formulated in the local system where the energy density and the particle number should be defined. To Lorentz-transform from the laboratory frame in the local rest frame one also needs to define a 4-velocity of the fluid element. If we know the components of the ideal stress tensor, $T_{\rm id}^{\mu\nu}=T^{\mu\nu}-\pi^{\mu\nu}$ and current $J^\mu=n\, u^\mu$, other quantities can be recovered as follows:
\begin{align}
n   &= J^0\,\sqrt{1-v^2}\,,\quad \epsilon=T_{\rm id}^{00}- M\,v\,,
\nonumber\\
M^2 &= T^{0x}_{\rm id}\,T^{0x}_{\rm id} + T^{0y}_{\rm id}\,T^{0y}_{\rm id} + T^{0z}_{\rm id}\,T^{0z}_{\rm id}\,.
\label{recover}
\end{align}
The modulus of the fluid velocity can be found as a root of the equation
\begin{align}
v=\frac{M}{T^{00}_{\rm id}-P\big(T^{00}_{\rm id}-M\, v,J^0\,\sqrt{1-v^2}\big)}
\label{recover-v}
\end{align}
and, therefore, depends on the chosen equation of state $P=P(\epsilon,n)$. The direction of the fluid velocity is determined as
\begin{align}
v^i=\frac{v}{M}\,T^{0i}_{\rm id}\,.
\label{recover-v-dir}
\end{align}

 We use in the code $\pi^{xy}$, $\pi^{xz}$, $\pi^{yz}$, $\pi^{yy}$, and $\pi^{zz}$ as independent variables. Other components can be recovered with the help of the following expressions
\begin{align}
 \pi^{00} =& \frac{1}{1-v_x^2}\Big[2 (\pi^{xy} v_x v_y+\pi^{xz} v_x v_z+\pi^{yz} v_y v_z)
\nonumber\\
&   -\pi^{yy}(v_x^2-v_y^2)
   -\pi^{zz}(v_x^2-v_z^2)\Big],
\nonumber\\
\pi^{0x} =& \frac{1}{1-v_x^2}
\Big[(\pi^{xy} v_y  +\pi^{xz}v_z )(1+v_x^2)
+2 \pi^{yz} v_x v_y v_z
\nonumber\\
& - \pi^{yy}v_x (1- v_y^2) - \pi^{zz}v_x (1- v_z^2 )
\Big],
\nonumber\\
\pi^{0y} =& \pi^{xy} v_x+\pi^{yy} v_y+\pi^{yz} v_z,
\nonumber\\
\pi^{0z} =& \pi^{xz} v_x+\pi^{yz} v_y+\pi^{zz} v_z,
\nonumber\\
\pi^{xx} =& \frac{1}{1-v_x^2}
\Big[2 (\pi^{xy} v_x v_y+\pi^{xz} v_x v_z+\pi^{yz} v_y v_z)
\nonumber\\
&-\pi^{yy}(1- v_y^2) -\pi^{zz}(1- v_z^2)
   \Big]\,.
   \label{recover-pi}
\end{align}
We emphasize that these expressions do not develop anomalously large values for the case of small fluid velocities.

\section{3$+$1 implementation of SHASTA algorithm}\label{app:shasta}
For completeness we provide the complete set of formulas for the 3+1 implementation of the SHASTA algorithm extending expressions provided in Ref~\cite{MNR2010}. The r.h.s. of Eq. (\ref{shastaform}) looks like
\begin{align}
    \vc{S}_{\rm cons}\!=\!\!
    \left[
    \begin{array}{l}
 0
 \\
 -\pd_t\pi^{00} - \Div(\vec{v} P) -
 (\pd_x\pi^{0x} + \pd_y\pi^{0y} + \pd_z\pi^{0z})
 \\
-\pd_t\pi^{0x} - \pd_xP - (\pd_x\pi^{xx}+\pd_y\pi^{xy}+\pd_z\pi^{xz})
\\
-\pd_t\pi^{0y} - \pd_yP - (\pd_x\pi^{yx} +\pd_y\pi^{yy} + \pd_z\pi^{yz})
\\
-\pd_t\pi^{0z} - \pd_zP - (\pd_x\pi^{zx} + \pd_y\pi^{zy}+\pd_z\pi^{zz})
    \end{array}
    \right].
\end{align}

For lattice realization of quantities $U(x,y,z,t)$ we will use notations
$U^{[n]}_{ijk}$, where index $n$ stands for temporal steps and $i,j,k$ for spatial lattice cells in $x,y$, and $z$ directions respectively.

At the first stage of the SHASTA algorithm one calculates, at each subsequent $(n +1)$th time step, the so-called transport-diffused solution
\begin{align}
\tilde U_{ijk}^{[n+1]} &= \tilde{\vU}_{ijk}^x + \tilde{\vU}_{ijk}^y + \tilde{\vU}_{ijk}^z - 2\vU_{ijk}^{[n]} + \Delta t\, {S}_{ijk},
\end{align}
\begin{widetext}
where $\vU_{ijk}^{[n]}$ is the full solution at the previous time step and
axillary quantities  $\tilde{\vU}_{ijk}^{x,y,z}$ are defined as
\begin{align}
\tilde{\vU}_{ijk}^x&=\frac12\left(\big[ Q_{ijk}^{x+} \big]^2
\big(\vU^{[n]}_{i+1,jk} - \vU^{[n]}_{ijk}\big)
- \big[Q_{ijk}^{x-}\big]^2
\big(\vU^{[n]}_{ijk} - \vU^{[n]}_{i-1,jk}\big) \right)
+(Q_{ijk}^{x+}+Q_{ijk}^{x-}) \, \vU^{[n]}_{ijk},
\\
\tilde{\vU}_{ijk}^y&=\frac12\left(
\big[Q_{ijk}^{y+}\big]^2 \big( \vU^{[n]}_{i,j+1,k}- \vU^{[n]}_{ijk} \big) - \big[Q_{ijk}^{y-}\big]^2 \big( \vU^{[n]}_{ijk}- \vU^{[n]}_{i,j-1,k} \big)
\right)
+(Q_{ijk}^{y+}+Q_{ijk}^{y-}) \, \vU^{[n]}_{ijk},
\\
\tilde{\vU}_{ijk}^z&=\frac12\left(\big[Q_{ijk}^{z+}\big]^2
\big(\vU^{[n]}_{ij,k+1} - \vU^{[n]}_{ijk}\big) - \big[Q_{ijk}^{z-}\big]^2\big(\vU^{[n]}_{ijk} - \vU^{[n]}_{ij,k-1}\big)\right)
+ (Q_{ijk}^{z+} + Q_{ijk}^{z-})\, \vU^{[n]}_{ijk}
\end{align}
with
\begin{align}
Q_{ijk}^{x\pm}&=\frac{1/2\mp \lambda\, (v_x)^{[n]}_{ijk}}
{1\pm\lambda \,\left[ (v_x)^{[n]}_{i\pm1,jk} - (v_x)^{[n]}_{ijk} \right]},\,
Q_{ijk}^{y\pm}=\frac{1/2\mp \lambda\, (v_y)^{[n]}_{ijk}}
{ 1\pm \lambda\, \left[(v_y)^{[n]}_{i,j\pm1,k} - (v_y)^{[n]}_{ijk}\right]},\,
Q_{ijk}^{z\pm}=\frac{1/2\mp \lambda\, (v_z)^{[n]}_{ijk}}
{1\pm \lambda \,\left[(v_z)^{[n]}_{ij,k\pm 1} - (v_z)^{[n]}_{ijk} \right]}.
\end{align}
The velocity components are taken at the $n$th time step.

Further, using the transport-diffused solution one calculates an antidiffusion flux that takes into account an anomalous diffusion
\begin{align}
A^{x,y,z}_{ijk}&=\frac{1}8 A_{\rm ad}^{x,y,z}\,\tilde{\vD}_{ijk}^{x,y,z},\quad
\tilde{\vD}_{ijk}^{x}=\tilde{\vU}_{i+1,jk}^x-\tilde{\vU}_{ijk}^x,\quad \tilde{\vD}_{ijk}^{y}=\tilde{\vU}_{i,j+1,k}^y-\tilde{\vU}_{ijk}^y,\quad
\tilde{\vD}_{ijk}^{z}=\tilde{\vU}_{ij,k+1}^z-\tilde{\vU}_{ijk}^z,
\end{align}
where $A_{\rm ad}^{x,y,z}$ are the antidiffusive mask coefficients. For simplicity, one takes them to be equal for all special directions and set $A_{\rm ad}=1$ as the default value.
Next, we calculate the limited antidiffusion fluxes
\begin{align}
\tilde{A}_{ijk}^x&=\sigma_{ijk}^x
\max\Big[0,\min\Big(\sigma_{ijk}^x\tilde{\vD}_{i+1,jk}^{x},\big|A^x_{ijk}\big|,
\sigma_{ijk}^x\tilde{\vD}_{i-1,jk}^{x}\Big)\Big],\nl
\tilde{A}_{ijk}^y&=\sigma_{ijk}^y
\max\Big[0,\min\Big(\sigma_{ijk}^y\tilde{\vD}_{i,j+1,k}^{y},\big|A^y_{ijk}\big|,
\sigma_{ijk}^y\tilde{\vD}_{i,j-1,k}^{y}\Big)\Big],\qquad
\sigma_{ijk}^{x,y,z}={\rm sgn} A^{x,y,z}_{ijk}.
\\
\tilde{A}_{ijk}^z&=\sigma_{ijk}^z
\max\Big[0,\min\Big(\sigma_{ijk}^z\tilde{\vD}_{ij,k+1}^{z},\big|A^z_{ijk}\big|,
\sigma_{ijk}^z\tilde{\vD}_{ij,k-1}^z\Big)\Big].\nonumber
\end{align}
The total incoming and outgoing antidiffusive fluxes in the cell are calculated as
\begin{align}
A^{\rm in}_{ijk}&=
 \max\big(0,\tilde A^x_{i-1,jk}\big)
-\min\big(0,\tilde A^x_{ijk}\big)
+\max\big(0,\tilde A^y_{i,j-1,k}\big)
-\min\big(0,\tilde A^y_{ijk}\big)
+\max\big(0,\tilde A^z_{ij,k-1}\big)
-\min\big(0,\tilde A^z_{ijk}\big),\\
A^{\rm out}_{ijk}&=
 \max\big(0,\tilde A^x_{ijk}\big)
-\min\big(0,\tilde A^x_{i-1,jk}\big)
+\max\big(0,\tilde A^y_{ijk}\big)
-\min\big(0,\tilde A^y_{i,j-1,k}\big)
+\max\big(0,\tilde A^z_{ijk}\big)
-\min\big(0,\tilde A^z_{ij,k-1}\big).
\end{align}
The maximal and minimal values of the transport-diffused solution $\vU^{[n+1]}_{ijk}$ after the antidiffusion stage are between
\begin{align}
\tilde{\vU}^{\min}_{ijk}&=
\min\left(\tilde{\vU}_{ij,k-1}^{[n+1]},
          \tilde{\vU}_{i,j-1,k}^{[n+1]},
          \tilde{\vU}_{i-1,jk}^{[n+1]},
          \tilde{\vU}_{ijk}^{[n+1]},
          \tilde{\vU}_{ij,k+1}^{[n+1]},
          \tilde{\vU}_{i,j+1,k}^{[n+1]},
          \tilde{\vU}_{i+1,jk}^{[n+1]}\right),\\
\tilde{\vU}^{\max}_{ijk}&=
\max\left(\tilde{\vU}_{ij,k-1}^{[n+1]},
          \tilde{\vU}_{i,j-1,k}^{[n+1]},
          \tilde{\vU}_{i-1,jk}^{[n+1]},
          \tilde{\vU}_{ijk}^{[n+1]},
          \tilde{\vU}_{ij,k+1}^{[n+1]},
          \tilde{\vU}_{i,j+1,k}^{[n+1]},
          \tilde{\vU}_{i+1,jk}^{[n+1]}\right).
\end{align}
This information is then used to determine the fractions of
the incoming and outgoing fluxes,
\begin{align}
F^{\rm in}_{ijk}  = \frac{1}{A^{\rm in}_{ijk}}
\big(\tilde{\vU}_{ijk}^{\max} - \tilde{\vU}_{ijk}^{[n+1]}\big),
\qquad
F^{\rm out}_{ijk} = \frac{1}{A^{\rm out}_{ijk}}
\big(\tilde{\vU}_{ijk}^{[n+1]} - \tilde{\vU}_{ijk}^{\min}\big).
\end{align}
The final antidiffusion fluxes are calculated as
\begin{align}
\hat A^x_{ijk}&=\tilde A^x_{ijk}
\Big[\min(1, F^{\rm in}_{i+1,jk},F^{\rm out}_{ijk})\Theta(\tilde A^x_{ijk})
+\min(1, F^{\rm in}_{ijk},F^{\rm out}_{i+1,jk})\Theta(-\tilde A^x_{ijk})\Big],\\
\hat A^y_{ijk}&=\tilde A^y_{ijk}
\Big[\min(1, F^{\rm in}_{i,j+1,k},F^{\rm out}_{ijk})\Theta(\tilde A^y_{ijk})
+\min(1, F^{\rm in}_{ijk},F^{\rm out}_{i,j+1,k})\Theta(-\tilde A^y_{ijk})\Big],\\
\hat A^z_{ijk}&=\tilde A^y_{ijk}
\Big[\min(1, F^{\rm in}_{ij,k+1},F^{\rm out}_{ijk})\Theta(\tilde A^z_{ijk})
+\min(1, F^{\rm in}_{ijk},F^{\rm out}_{ij,k+1})\Theta(-\tilde A^z_{ijk})\Big].
\end{align}
Finally, the full solution for $n+1$ time step is given by
\begin{align}
\vU_{ijk}^{[n+1]}&=\tilde U_{ijk}^{[n+1]}
+\big(\hat A^x_{i-1,jk}-\hat A^x_{ijk}\big)
+\big(\hat A^y_{i,j-1,k}-\hat A^y_{ijk}\big)
+\big(\hat A^z_{ij,k-1}-\hat A^z_{ijk}\big).
\end{align}

\end{widetext}

\end{document}